\documentclass[aps,prc,superscriptaddress,showpacs,showkeys]{revtex4}
\usepackage{epsfig,pstricks,dcolumn}

\begin{document}

%\preprint{}

\title{Composition and thermodynamics of nuclear matter with light clusters}

\author{S. Typel}
%\email[]{s.typel@gsi.de}
%\thanks{}
\email{s.typel@gsi.de}
\affiliation{Excellence Cluster Universe, 
Technische Universit\"{a}t M\"{u}nchen, Boltzmannstra\ss{}e 2, D-85748
Garching, Germany}
\affiliation{GSI Helmholtzzentrum f\"{u}r Schwerionenforschung GmbH, 
Theorie, Planckstra\ss{}e 1, D-64291 Darmstadt, Germany}

\author{G. R\"{o}pke}
\email{gerd.roepke@uni-rostock.de}
\affiliation{Institut f\"{u}r Physik, Universit\"{a}t Rostock,
Universit\"{a}tsplatz 3, D-18051 Rostock, Germany}

\author{T. Kl\"{a}hn}
\email{thomas.klaehn@googlemail.com}
\affiliation{Theory Group, Physics Division, Building 203,
Argonne National Laboratory,
9700 South Cass Avenue, Argonne, IL 60439, USA}
\affiliation{Instytut Fizyki Teoretycznej, Uniwersytet Wroc\l{}awski,
Max Born Place 9, 50-204 Wroc\l{}aw, Poland}

\author{D. Blaschke}
\email{blaschke@ift.uni.wroc.pl}
\affiliation{Instytut Fizyki Teoretycznej, Uniwersytet Wroc\l{}awski,
Max Born Place 9, 50-204 Wroc\l{}aw, Poland}
\affiliation{Bogoliubov Laboratory for Theoretical Physics, JINR Dubna,
Joliot-Curie Street 6,  141980 Dubna, Russia}

\author{H.H. Wolter}
\email{hermann.wolter@physik.uni-muenchen.de}
\affiliation{Fakult\"at f\"ur Physik, Universit\"{a}t M\"{u}nchen,
Am Coulombwall 1, D-85748 Garching, Germany}


%\date{\today}

% \begin{abstract}
% % insert abstract here
% \end{abstract}

%\pacs{21.60.-n, 21.30.Fe, 21.65.+f, 21.10.-k}
%\keywords{}

%\maketitle

%%%%%%%%%%%%%%%%%%%%%%%%%%%%%%%%%%%%%%%%%%%%%%%%%%%%%%%%%%%%%%%%%%%%%%%%%%%%

%\section{}
%\subsection{}
%\subsubsection{}

%\section{Introduction}

\begin{abstract}
We investigate nuclear matter at finite temperature  and density, 
including the formation of light clusters up to the $\alpha$ particle 
($1<A<4$).
The novel feature of this work is to include the formation of 
clusters as well as their dissolution due to medium effects in a 
systematic way using two many-body theories: a microscopic quantum 
statistical (QS) approach and a generalized relativistic mean field (RMF) 
model. 
Nucleons and clusters are modified by medium effects.
While the nucleon quasiparticle properties are determined within the RMF model
from the scalar and vector self energies, the cluster binding energies
are reduced due to Pauli blocking shifts calculated in the QS approach. 
Both approaches reproduce the limiting cases of
nuclear statistical equilibrium (NSE) at low densities and cluster-free 
nuclear matter at high densities. 
The treatment of the cluster dissociation is based on the Mott effect due to
Pauli blocking, implemented in slightly different ways in the QS and the 
generalized RMF approaches. 
This leads to somewhat different results in the intermediate density 
range of about $10^{-3}$ to $10^{-1}$~fm$^{-3}$, which gives an estimate of 
the present accuracy of the theoretical predictions.
We compare the numerical results of these models for cluster 
abundances and thermodynamics in the region of medium excitation energies with 
temperatures $T \le 20~{\rm MeV}$ and baryon number densities from zero to a 
few times saturation density.
The effect of cluster formation on the liquid-gas phase transition
and on the density dependence of the symmetry energy is studied.
It is demonstrated that the parabolic approximation for the asymmetry 
dependence of the nuclear equation of state breaks down at low temperatures 
and at subsaturation densities due to cluster formation.
Comparison is made with other theoretical approaches, in particular those, 
which are commonly used in astrophysical calculations.
The results are relevant for heavy-ion collisions and astrophysical 
applications.
\end{abstract}

\date{\today}
\maketitle

\section{Introduction}

The composition and the equation of state (EoS) of nuclear matter, 
possible phase transitions or condensates are widely discussed
in several areas of nuclear physics. They attain an 
increasing importance in astrophysics and cosmology.
Many of the systems under consideration have large charge asymmetries
and cover a broad range of densities.
Therefore, a deeper understanding of the composition and the thermodynamical 
properties of nuclear matter, and, in particular, the density dependence of 
the symmetry energy, from very low to super-saturation densities is of 
great interest. 
Let us mention three examples for systems where the knowledge of the symmetry 
energy in a broad range of densities is of crucial importance:
(1) the surface structure of 
exotic nuclei with large neutron excess or new exotic collective modes,
(2) the structure and composition of neutron stars
from the ultra-dense core to the crust at subsaturation densities with varying 
asymmetry due to $\beta$-equilibrium, and  
(3) core collapse supernovae, where at high 
densities the symmetry energy
determines the energy of the shock, and at low densities it 
affects the nuclear composition, neutrino interactions and aspects of 
nucleosynthesis. 

This wide spectrum of applications illustrates the importance of gathering 
reliable information on the EoS of nuclear matter, and, specificially, the 
symmetry energy, since predictions of its density dependence differ strongly 
for different theoretical approaches. 
Therefore, considerable efforts have been taken to obtain constraints from
observational data on all aspects of the EoS.
A promising source of information arises from the study of heavy-ion 
collisions, where transient states of very different densities can be 
investigated and the asymmetry can be varied to a certain extent by choosing 
the collision system.
Central collisions at high energies yield large compressions and provide
information from observations of nucleon flow \cite{Danielewicz:2002pu} and 
particle production as shown for kaons, e.g., in 
\cite{Fuchs:2000kp,Ferini:2006je}. 
Recent reviews of studies of the symmetry energy in heavy-ion collisions were 
given in \cite{Baran:2004ih,Steiner:2004fi,Li:2008gp}.

Recently, particular interest has been devoted to the properties of nuclear 
and neutron matter at very low densities, down to the limit of zero density. 
Below saturation density, correlations are expected to become important 
and nuclear matter can become inhomogeneous.
An aspect of this is seen in fragmentation as a signature of the liquid-gas 
phase transition. 
At even smaller densities, down to one hundredth or one thousandth of 
saturation density and at moderate temperatures, few-body correlations remain 
important. 
This results from the fact, that at low densities the system can 
minimize its energy by forming light clusters such as deuterons, or 
particularly strongly bound $\alpha$ particles.
With increasing density such clusters will dissolve due to the Pauli principle.
Thus there exists an interesting detailed evolution of the correlations and 
the composition  in nuclear matter depending on the density and temperature, 
which presents a challenge to a theoretical description.

In the laboratory, low-density matter occurs in the outer regions of heavy 
nuclei, in halo nuclei, in
expanding hot matter from heavy-ion reactions, in the envelopes 
of core- collapse supernovae, and also in recently
discussed low-density isomeres such as the Hoyle state of $^{12}$C 
\cite{Tohsaki:2001an}.
Experimental information on the composition and behavior of very low density 
nuclear matter was recently obtained by Kowalski et al.\
\cite{Kowalski:2006ju} from the observation of the light particles emitted in 
low-energy heavy-ion collisions. 
It was found that the fraction of light particles is substantial at very low 
densities, implying a finite symmetry energy in the limit of zero density.

Nonrelativistic Skyrme Hartree-Fock and relativistic mean-field calculations 
revealed a tight correlation between the density derivative of the neutron
matter EoS near 2/3 of the saturation density and the neutron skin thickness
$\Delta R_{np}$ of heavy nuclei \cite{Bro00,Typ01}. 
This observation translates into a correlation of the density dependence of 
the symmetry energy at saturation density with $\Delta R_{np}$
\cite{Li:2008gp,Che05,Fur02,Bal04,Ava07,Cen09,War09}.
The scheduled PREX experiment at JLab is expected to provide a precise
value of the neutron skin thickness of ${}^{208}$Pb from the
observation of parity-violations in electron scattering \cite{Hor01},
hence providing an independent constraint on the density dependence
of the symmetry energy.

At densities beyond nuclear saturation constraints on the EoS are expected in 
increasing quality from neutron star observables as, e.g., masses, mass-radius 
relations, gravitational binding energy and the cooling behavior of neutron 
stars.
A recent review by Kl\"ahn et al. \cite{Klahn:2006ir} discusses these 
constraints in confrontation with those from heavy-ion collisions.
It demonstates that at present it is far from trivial to obtain an EoS 
which is consistent with all these available observational data.
The study of exotic structures  like pasta phases in the inner crust 
of neutron stars gives information on very neutron rich matter around normal 
densities 
\cite{Avancini:2008kg,Horowitz:2008vf,Sonoda:2007ni,Maruyama:2006jk,Newton:2007bk}.

The actual composition of very low density matter is also relevant
for the investigation of various stages in supernova explosions
as pointed out in a number of recent publications.
It is known to affect the effectiveness of the neutrino reheating
of the shock wave \cite{Lattimer:2004pg}.
Also, the  stellar core collapse is mostly determined 
during the dynamical plunge phase at 
densities between $10^{12}$ and $10^{14}$ g/cm$^3$, where temperatures between
$10^{10}$ and $10^{11}$ K are reached 
\cite{Dimmelmeier:2007ui,Dimmelmeier:2008iq}. 
The sensitivity of the collapse dynamics on the properties of 
matter in this density regime could strongly influence the 
structure and composition of the proto-neutron star \cite{Sumiyoshi:2008qv}
and possible gravitational wave and neutrino signals 
emitted in various stages during and after core bounce.
 
Only very few models for the EoS are applicable in actual supernova 
simulations. 
The reason is found in the required wide range of temperatures, 
densities and asymmetries, which are either not available, e.g.\ in tabular
form, or not covered by the model.
The most frequently used EoS are those of Lattimer and Swesty 
\cite{Lattimer:1991nc} and of Shen, Toki et al.\ \cite{Shen:1998gq}.
The former one is based on an extended liquid drop model 
for the nuclei embedded in a nucleon and 
$\alpha$-particle gas,
while the latter one has been developed in the framework of the Relativistic 
Mean Field (RMF) approach with non-linear meson self-couplings 
\cite{Serot:1984ey}. 
Shen et al. use the Thomas-Fermi approximation to describe heavy nuclei 
embedded in a gas. 
They consider $\alpha$ particles as a separate species, using an excluded
volume prescription to model the dissolution of $\alpha$ particles
at high densities which accounts for medium effects only in a very global way. 
The model neglects other light clusters such as deuterons, tritons, and 
helions (${}^{3}$He).

The problem of cluster formation in low-density nuclear matter has been 
addressed recently based on a virial expansion \cite{Horowitz:2006pj} 
to obtain the EoS for nuclear matter at all asymmetries including nucleons
and $\alpha$ particles (${}^{4}$He) \cite{Horowitz:2005nd} and in later 
work also tritons and helions \cite{O'Connor:2007eb} (see also \cite{Mal08} 
for a closely related formulation in an S-matrix approach
and the quasiparticle gas model \cite{Heckel:2009br}).
The virial coefficients in the Beth-Uhlenbeck approach 
\cite{Beth:1937} are given by the cluster bound state energies
and scattering phase shifts.
In Ref.\ \cite{Horowitz:2005nd} 
these were taken directly from experiment, thus providing an exact limit for 
the EoS at very low densities, where the scattering is not yet influenced by 
medium effects. 
These results are believed to be reliable for densities up to 
$n_{\rm sat}/1000$ and not too small temperatures.
They provide a benchmark for other calculations.

The present paper emphasizes that correlations, in-medium
modifications of cluster properties and mean-field effects have to be
considered simultaneously in the description of low-density nuclear 
matter, since all these affect the thermodynamical properties of the EoS. 
The occurence of clusters also changes the symmetry energy, since the cluster 
correlations depend on the asymmetry of the system.
Here we will restrict ourselves to matter in thermodynamic 
equilibrium at temperatures
$T \leq 20$ MeV and baryon number densities $n \leq $ 0.2 fm$^{-3}$,
where the quark substructure and excitations of internal degrees of
freedom of nucleons (protons and neutrons) are not important
and the nucleon-nucleon interaction can be represented by an
effective interaction potential.

In this work, we explore two approaches to the problem: one is a 
quantum statistical (QS) formulation based on the thermodynamic 
Green function method \cite{Ropke:1982}. This approach makes 
explicit use of an effective nucleon-nucleon interaction. 
It allows us to account for medium effects on the cluster
properties. The second approach is a generalized relativistic 
mean-field (RMF) model, where the medium modified 
clusters are introduced as explicit degrees of freedom. 
The two methods have their strengths and deficiencies. The RMF method
is reliable to determine the nucleon quasi-particle propagator
in the medium, and this information is introduced into the QS model. 
The QS model, on the other hand, can determine the medium 
modifications of the clusters, such as the mass shift and the 
(momentum-dependent) Mott densities, where the clusters get dissolved. 
These are then introduced in parametrized form into the RMF method.

The particle number density $n_{\tau} (T, \mu_{p}, \mu_{n})$ 
of protons ($\tau = p$) or neutrons ($\tau = n$) depends on
the temperature $T$ and the chemical 
potentials $\mu_{\tau}$. In the QS approach  
it is obtained from the
single-particle spectral function, which can be expressed in terms of the
self-energy. This is the main quantity to be evaluated.
Considering the ladder approximation \cite{Ropke:1982,Ropke:1983}, 
the formation of bound states is taken into account 
in a Bethe-Goldstone equation which
in the low-density limit reduces to an effective Schr\"odinger equation.
Effects of the medium can be included in a self-consistent way within
the cluster mean-field approximation (for references see 
\cite{Ropke:1980,Ropke:1983,Dukelsky:1998}).
The bound state energies are also modified due to Pauli blocking in
the correlated medium.
An extended discussion of the two-particle problem can be found in 
\cite{Schmidt:1990}.
This generalized Beth-Uhlenbeck formulation accounts for
medium effects suppressing correlations at high densities. 
It allows to determine the second virial 
coefficient and also contains the Brueckner approach to matter near saturation
density. 
The approach has
been extended to three and four-particle bound states in Refs.
\cite{Ropke:1983,Ropke:1982a}. The medium dependent shift of the 
cluster binding energies has been investigated in 
\cite{Beyer:1999zx,Sedrakian:2005db}.  
We emphasize again that 
this quantum statistical approach avoids the introduction of 
semi-empirical
concepts such as the excluded volume mechanism to mimic in-medium
effects. However, since the quasi-particle propagator is introduced 
from the outside (i.e. from the RMF model) the back effect of the clusters 
on the mean field is not included.

The RMF model, on the other hand, takes this back reaction fully into 
account. 
On the other hand, only the bound state contributions of the clusters are 
included 
(from the QS model). Thus the continuum contributions are missing,
which, as we shall see, leads to an overestimation of the two-particle 
correlations.
Thus, by investigating both these models we also obtain an estimate 
of the remaining uncertainties of theoretical approaches. Also, in the 
present work, in both approaches, we have not yet included the contribution
of heavier clusters, that should appear in the intermediate density range 
before the matter becomes homogenous again at densities near saturation. 
This problem will be treated in a later work.

Extending the quasiparticle approach including the formation of light 
clusters enables us to describe the smooth transition from the 
low-density limit, where the nuclear statistical equilibrium (NSE) 
or the virial expansion are
applicable, to the region of the saturation density where mean-field concepts 
have been successfully applied. 
None of the existing approaches to model the EoS gives satisfactory
results in both regions simultaneously.
More precisely, 
the EoS of Lattimer and Swesty \cite{Lattimer:1991nc} as well as 
the EoS of Shen, Toki et al. \cite{Shen:1998gq} fail to reproduce the NSE in 
the low-density limit, while the EoS in the virial expansion 
\cite{Horowitz:2006pj,Horowitz:2005nd,O'Connor:2007eb} 
ignores medium effects on light clusters and cannot describe the
dissolution of clusters at high densities.

The outline of this paper is as follows: In Sect.\ \ref{sec:QSA} 
we review the QS approach to the EoS and put particular emphasis 
on the calculation of the 
medium modifications of the clusters. 
As one result we obtain a density and temperature dependent 
modification of 
the binding energies of the clusters. In Sect.\ \ref{sec:RMF} 
we introduce our generalized RMF model with light clusters as 
explicit degrees of freedom. 
We use a RMF model with density dependent meson-nucleon couplings
\cite{Typel:2005ba}, which was used very successfully to describe nuclear 
structure in a wide region of the nuclear chart and has also been tested in 
heavy-ion collisions. 
The medium dependent masses of the clusters lead to a coupling of the nucleon 
and cluster dynamics. 
We also show how the thermodynamic quantities, such as free and internal 
energy, pressure and entropy, are obtained as functions of density and 
temperature. 
In Sect.\ \ref{sec:SNM} we discuss the composition of nuclear matter
and present the thermodynamical quantities 
for symmetric nuclear matter in both approaches. 
We also compare to the NSE model which
gives the correct low density limit.
Of particular interest is the phase transition from (partially) clusterized to 
the homogeneous medium which is considered in Sect.\ \ref{sec:pt}.
In Sect.\ \ref{sec:SymE} we discuss specifically the symmetry energy of nuclear
matter as a function of density and temperature, 
which is drastically 
changed at very low densities due to the cluster correlations.
In Sect.\ \ref{sec:alpha} 
we finally compare the results for the $\alpha$ cluster fraction
in the QS and generalized RMF models
with previous approaches and 
discuss the advantages of the present ones. 
We close with an outlook on further work, which should finally lead to an EoS
that can be used in a wide range of problems, including nuclear structure, 
heavy-ion reactions, and supernovae simulations.
Throughout the paper we use natural units where 
$\hbar=c=k_{B}=1$.

\section{Quantum statistical approach to the equation of state}
\label{sec:QSA}

\subsection{Single-particle spectral function and quasiparticles} 
\label{sec2.1}

Using the finite-temperature Green function formalism, 
a non-relativistic quantum statistical approach 
can be given to describe the equation of state of nuclear matter
including the formation of bound states \cite{Ropke:1983,Schmidt:1990}.
It is most convenient to start with the nucleon number densities 
$n_{\tau}(T,\tilde{\mu}_{p},\tilde{\mu}_{n})$ as
functions of temperature 
$T$ and non-relativistic 
chemical potentials $\tilde{\mu}_{\tau}$ 
for protons ($\tau = p$) and neutrons ($\tau = n$), respectively,
\begin{equation}
 n_{\tau}(T,\tilde{\mu}_p,\tilde{\mu}_n)=
 \frac{1}{\Omega} \sum_{1} \langle a^{\dagger}_{1} a_{1} \rangle 
 \delta_{\tau,\tau_{1}} 
 = 2  \int \frac{d^{3}k_{1}}{(2 \pi)^{3}}
 \int_{-\infty}^\infty \frac{d \omega}{2 \pi} f_{1,Z}(\omega) 
 S_{1}(1,\omega)\,, 
\label{eosspec}
\end{equation}
where $\Omega$ is the system volume, $\{1\}=\{ k_{1}, 
\sigma_{1}, \tau_{1}\}$ 
denotes the single-nucleon quantum numbers momentum, spin, and isospin. 
Summation over spin yields the factor $2$ and 
\begin{equation}
\label{faz}
 f_{A,Z}(\omega) = \left(\exp\left\{ 
\beta \left[\omega -Z \tilde{\mu}_{p} -(A-Z) \tilde{\mu}_{n}\right] 
 \right\} - (-1)^{A}\right)^{-1}
\end{equation}
is the Fermi or Bose distribution function which depends on the inverse 
temperature $\beta = 1/T$. The non-relativistic chemical
potential $\tilde{\mu}_{\tau}$ is related to the relativistic chemical
potential $\mu_{\tau}$ by
$\mu_{\tau} = \tilde{\mu}_{\tau}+m_{\tau}$ with the nucleon mass
$m_{\tau}$.
Instead of the isospin quantum number $\tau$ 
we occasionally use the mass number $A$ and the charge number $Z$.
Both the distribution function and the spectral function $S_1(1,\omega)$ 
depend on the temperature and the chemical potentials $\tilde{\mu}_{p}$, 
$\tilde{\mu}_{n}$, not given explicitly. 
We work with a grand canonical ensemble and have to invert Eq. (\ref{eosspec})
to write the chemical potentials as functions of the 
densities $n_{p}$, $n_{n}$. 
For this EoS, expressions such as the Beth-Uhlenbeck formula and 
its generalizations have been derived 
\cite{Ropke:1983,Schmidt:1990,Horowitz:2005nd}.

We consider both the total number densities of protons and neutrons,
$n_{p}^{\rm tot}$ and $n_{n}^{\rm tot}$, and the temperature $T$
as given parameters.
Alternatively, the total 
baryon density $n=n_{n}^{\rm tot}+n_{p}^{\rm tot}$ 
and the asymmetry of nuclear matter 
$\delta = (n_{n}^{\rm tot}-n_{p}^{\rm tot})/n=1-2Y_{p}$ are used.
$Y_{p}$ denotes the total proton fraction.
In addition to the frozen equilibrium where $n_{p}^{\rm tot}$ and
$n_{n}^{\rm tot}$ are given, we assume 
homogeneity and isotropy in space. Thermodynamical stability is considered in
Sects. \ref{sec:pt} and \ref{subsec:tq}.
In a further development, allowing for weak interactions, 
$\beta$-equilibrium may
be considered, which is of interest for astrophysical applications. 
In that case the asymmetry $\delta$ is uniquely determined for given
$n$ and $T$.

The spectral function $S_{1}(1,\omega)$ 
is related to the self-energy $\Sigma(1,z)$ according to
\begin{equation}
\label{spectral}
 S_{1}(1,\omega) = \frac{2 {\rm Im}\,\Sigma(1,\omega-i0)} 
{[\omega - E(1)- {\rm Re}\, \Sigma(1,\omega)]^2 + 
[{\rm Im}\, \Sigma(1,\omega-i0)]^2 } \: ,
\end{equation}
where the imaginary part has to be taken for a small negative imaginary part 
in the frequency $\omega$. $E(1)=k_{1}^{2}/(2m_{1})$ is the
kinetic energy of the free nucleon.
The solution of the relation 
\begin{equation}
\label{quasinucleon}
E_{1}^{\rm qu}(1) = E(1) + {\rm Re}\, \Sigma[1,E_{1}^{\rm qu}(1)] 
\end{equation}
defines the single-nucleon quasiparticle energies 
$E_{1}^{\rm qu}(1) = E(1) + \Delta E^{\rm SE}(1)$. 
Expanding for small Im~$\Sigma(1,z)$, the spectral function yields a 
$\delta$-like contribution. The densities
are calculated from Fermi
distributions with the quasiparticle energies so that
\begin{equation}
 n^{\rm qu}_{\tau}(T,\tilde{\mu}_{p},\tilde{\mu}_{n})
 =\frac{2}{\Omega} \sum_{k_{1}} f_{1,Z}[E_{1}^{\rm qu}(1)] 
\label{nqu}
\end{equation}
follows for the EoS in mean field approximation.
This result does not contain the contribution of bound states and therefore 
fails to be correct in the low-temperature, low-density limit where the NSE 
describes the nuclear matter EoS.

As shown in Refs.\ 
\cite{Ropke:1983,Schmidt:1990}, the bound state contributions
are obtained from the poles of Im~$\Sigma(1,z)$ which cannot be neglected in 
expanding the spectral function with respect to Im~$\Sigma(1,z)$. 
A cluster decomposition of the self-energy has been proposed, see 
\cite{Ropke:1983}.
The self-energy is expressed in terms of the $A$-particle Green functions which
read in bilinear expansion
\begin{equation}
 G_A(1...A,1^{\prime}\dots A^{\prime},z_A)
  =\sum_{\nu K} \psi_{A \nu K}(1\dots A)
 \frac{1}{z_{A}-E^{\rm qu}_{A,\nu}(K)} 
 \psi^{\ast}_{A \nu K}(1^{\prime}\dots A^{\prime}) \: .
\label{bilinear}
\end{equation}
The $A$-particle wave function $\psi_{A \nu K}(1\dots A) $
and the corresponding eigenvalues $E^{\rm qu}_{A, \nu}(K)$ result from solving 
the in-medium Schr\"{o}dinger equation (see the following subsections).
$K$ denotes the center of mass momentum of the $A$-nucleon system. 
Besides the bound states, the summation over the internal quantum states $\nu$ 
includes also the scattering states.

The evaluation of the equation of state in the low-density limit is
straightforward.
Considering only the bound-state contributions, we obtain the result 
\begin{eqnarray}
\label{EoS:p}
 n^{\rm tot}_{p}(T,\tilde{\mu}_p,\tilde{\mu}_n)&=& 
 \frac{1 }{\Omega} \sum_{A,\nu,K}Z 
 f_{A,Z}[E^{\rm qu}_{A,\nu}(K;T,\tilde{\mu}_p,\tilde{\mu}_n)]\,,
 \nonumber \\  
 n^{\rm tot}_{n}(T,\tilde{\mu}_p,\tilde{\mu}_n)&=& 
 \frac{1}{\Omega} \sum_{A,\nu,K}(A-Z) 
 f_{A,Z}[E^{\rm qu}_{A,\nu}(K;T,\tilde{\mu}_p,\tilde{\mu}_n)]\,
\label{quasigas}
\end{eqnarray}
for the EoS describing a mixture of components (cluster quasiparticles) obeying
Fermi or Bose statistics. 
The total baryon density results as 
$n(T,\tilde{\mu}_{p},\tilde{\mu}_{n})
= n^{\rm tot}_{n}(T,\tilde{\mu}_{p},\tilde{\mu}_{n})
 +n^{\rm tot}_{p}(T,\tilde{\mu}_{p},\tilde{\mu}_{n})$. 
To derive the extended Beth-Uhlenbeck formula, see \cite{Ropke:1982a},
we restrict the
summation to $A \le 2$, but extend the summation over the internal quantum
numbers $\nu$,
not only to the excited states, but also the scattering states.
Note that at low temperatures Bose-Einstein condensation may occur.

The NSE is obtained in the low-density limit if the in-medium energies  
$E^{\rm qu}_{A,\nu}(K;T,\tilde{\mu}_{p},\tilde{\mu}_{n})$ 
can be replaced by the binding energies of the isolated nuclei 
$E^{(0)}_{A,\nu}(K)=E_{A,\nu}^{(0)}+K^{2}/(2Am)$, with
$m=939$~MeV the average nucleon mass. 
For the cluster contributions, i.e.\  $A>1$, the summation
over the internal quantum numbers is again restricted to the bound states only.
We have
\begin{eqnarray}
\label{NSE:p}
 n^{\rm NSE}_{p}(T,\tilde{\mu}_p,\tilde{\mu}_n)&=& 
 \frac{1 }{\Omega} \sum_{A,\nu,K}^{\rm bound} Z 
 f_{A,Z}[E^{(0)}_{A,\nu}(K)]\,,
 \nonumber \\  
 n^{\rm NSE}_{n}(T,\tilde{\mu}_p,\tilde{\mu}_n)&=& 
 \frac{1}{\Omega} \sum_{A,\nu,K}^{\rm bound} (A-Z) 
 f_{A,Z}[E^{(0)}_{A,\nu}(K)]\,.
\end{eqnarray}
The summation over $A$ includes also the contribution of
free nucleons, $A=1$.

In the nondegenerate and nonrelativistic case assuming a
Maxwell-Boltzmann distribution, the summation over the 
momenta $K$ can be performed analytically and the
thermal wavelength 
$\lambda= \sqrt{2 \pi/(m T)}$ of the nucleon enters. 
As shown below, the medium effects in nuclear matter are negligible
below 10$^{-4}$ times the saturation density $n_{\rm sat}$ 
for the temperatures considered here.

Interesting quantities are the mass fractions
\begin{equation}
 X_{A,Z}= \frac{A}{\Omega n} \sum_{\nu, K}
 f_{A,Z}[E^{\rm qu}_{A,\nu}(K;T,\tilde{\mu}_p,\tilde{\mu}_n)]
\end{equation}
of the different clusters. From the EoS considered here,
thermodynamical potentials can be obtained by integration,
in particular the free energy per volume $F/\Omega$. In the special case of 
symmetric nuclear matter,
$Y_p^{\rm s}=0.5$, the free energy per volume is obtained 
from the averaged chemical potential 
$\tilde \mu =(\tilde \mu_p+\tilde \mu_n)/2$ as
\begin{equation}
\label{eq:freV}
 F(T,n,Y_p^{\rm s}) / \Omega  = \int_0^n dn' \:
 \tilde \mu(T,n',Y_p^{\rm s}) \:  .
\end{equation}

In the quantum statistical approach described above,
we relate the EoS to properties of the correlation functions, in particular to
the peaks occurring in the $A$-nucleon spectral function describing the 
single-nucleon quasiparticle ($A=1$) as well as the nuclear quasiparticles 
($A \ge 2$).  
Different approaches to these quasiparticle energies can be given 
by calculating self-energies
which reproduce known properties of the nucleonic system. 
In the following Subsects. \ref{subsec:medium1} and 
\ref{subsec:medium2} we discuss results obtained 
from a microscopic Hamiltonian approach to nuclear matter. 
In Sect.\ \ref{sec:RMF} 
a relativistic mean field (RMF) approach is given, which 
is based on an effective nucleon-meson Lagrangian.

\subsection{Medium modification of single nucleon properties}
\label{subsec:medium1}

The single-particle spectral function contains the
single-nucleon quasiparticle contribution,
$E_{1}^{\rm qu}(1) = E^{\rm qu}_{\tau}(k)$, given in Eq.\ 
(\ref{quasinucleon}), 
where $\tau$ denotes isospin of particle $1$ and $ k$ is the momentum.
In the effective mass approximation, the single-nucleon quasiparticle
dispersion relation reads
\begin{equation}
\label{quasinucleonshift}
E_{\tau}^{\rm qu}(k) = \Delta E^{\rm SE}_{\tau}(0) +\frac{k^2}{2
m_\tau^{\ast}} + {\mathcal O}(k^4)\,,
\end{equation}
where the quasiparticle energies are shifted at zero momentum $ k$
by $\Delta E^{\rm SE}_{\tau}(0)$, and $m_\tau^{\ast}$ 
denotes the effective mass of neutrons
($\tau=n$) or protons ($\tau=p$).
Both quantities, $\Delta E^{\rm SE}_{\tau}(0)$ and $m_\tau^{\ast}$, are
functions of $T$, $n_{p}$ and $n_{n}$, characterizing the surrounding matter.

Expressions for the  single-nucleon quasiparticle energy 
$E^{\rm qu}_{\tau}(k)$ can be given by the Skyrme parametrization
\cite{Vautherin:1971aw} or by more sophisticated approaches such as 
relativistic mean-field approaches \cite{Serot:1984ey}, 
see Sec.\ \ref{sec:RMF}, 
and relativistic Dirac-Brueckner Hartree-Fock \cite{Margueron:2007jc} 
calculations.
We will use the density-dependent relativistic mean field approach of
 \cite{Typel:2005ba} that is designed not only to reproduce known 
properties of nuclei,
but also agrees with microscopic calculations in the low density
region. It is expected that this approach gives at present an 
optimal fit to the quasiparticle energies 
and is applicable in a large interval of densities and temperatures.
 
Microscopic calculations are based on a model describing the
interaction between the nucleons. To go beyond the mean-field
approximation, strong interaction as well as bound state formation has
to be taken into account. This can be done in the low-density region 
where in the non-relativistic case a T-matrix can be introduced. 
We start from a nonrelativistic Hamiltonian in fermion second quantization
\begin{equation}
\label{Ham}
 H = \sum_1 E(1)a_{1}^{\dagger} a_{1} +\frac{1}{2} 
\sum_{12,1^{\prime}2^{\prime}} V(12,1^{\prime}2^{\prime}) 
 a_{1}^{\dagger} a_{2}^{\dagger} a_{2^{\prime}} a_{1^{\prime}} \: ,
\end{equation}
where the kinetic energy 
is $E(1)=P_{1}^{2} / (2 m_{1})$, and the potential energy contains 
the matrix element $V(12,1^{\prime}2^{\prime})$ 
of the nucleon-nucleon interaction.

Since there is no fundamental expression for the nucleon-nucleon interaction, 
a phenomenological form is assumed 
to reproduce empirical data such as the nucleon scattering phase shifts. 
Different parametrizations are in use. 
For calculations one can use potentials such as PARIS and BONN or their 
separable representations \cite{Haidenbauer:1984dz}. 
To obtain the empirical parameter values of nuclear matter at saturation 
density, three-body forces have been introduced in the Hamiltonian (\ref{Ham}).
In particular, the Argonne AV18/UIX potential \cite{Wiringa:1994wb} has been 
used to calculate light nuclei \cite{Pieper:2001mp}.

Replacing the two-particle $T$-matrix in Born approximation with the 
interaction potential $V$, we obtain the Hartree-Fock approximation for the 
energy shift 
\begin{equation}
\Delta E^{\rm HF}(1) = \sum_{2}[V(12,12)-V(12,21)] 
 f_{1,\tau_{2}}[E(2)+\Delta E^{\rm HF}(2)] \: .
\end{equation}
In this approximation, all correlations in the medium are neglected. 
The self-energy does not depend on frequency, i.e.\ it is instantaneous 
in time, with vanishing imaginary part.

A full Dirac-Brueckner Hartree-Fock (DBHF)
calculation has been performed by Fuchs \cite{Fuchs:2005yn}
and has been compared with RMF approaches.
The relation between the $T$ matrix approach 
and the Brueckner $G$ matrix approach
was discussed in detail in Ref.\ \cite{Alm:1996}. 
Extended work has been performed using sophisticated interaction potentials  
to evaluate the quasiparticle energies in the DBHF approximation, for recent 
reviews see Refs.\ \cite{Fuchs:2005yn,vanDalen:2005ns,Margueron:2007jc}.
There was reasonable agreement between the RMF parametrisation of the 
quasiparticle energies and the DBHF results. 

We can assume \cite{Klahn:2006ir} that 
the density-dependent RMF parametrisation covers a large density 
region {(which will be discussed in detail in Sect.\ \ref{sec:RMF})} 
and that it can be used instead of the above Hartree-Fock shifts to determine 
the single-nucleon quasiparticle energies. 
They result as
\begin{equation}
E^{\rm qu}_{n,p}(0)=\sqrt{[m - \Sigma_{n,p}(T,n,\delta)]^2 + k^2} 
			+ \Sigma^0_{n,p}(T,n,\delta) \: , 
\end{equation}
where $\Sigma_{n,p}$ and $\Sigma^0_{n,p}$ are the scalar  
and the time component of the vector self energy, respectively.
%for neutrons and
%\begin{equation}
%E^{\rm qu}_{p}(k)=\sqrt{[m^2 - 
%                 S(T,n,-\delta)]^2 + k^2} + V(T,n,-\delta)
%\end{equation}
%for protons.
In the nonrelativistic limit, the shifts of the quasiparticle energies are
\begin{equation}
\Delta E^{\rm
     SE}_{n,p}(k)=\Sigma^0_{n,p}(T,n,\delta)-\Sigma_{n,p}(T,n,\delta)\: .
\end{equation}
%and
%\begin{equation}
%\Delta E^{\rm qu}_{p}(k)=V(T,n,-\delta)-S(T,n,-\delta) \: .
%\end{equation}
The effective masses for neutrons and protons are given by
\begin{equation}
m_{n,p}^{\ast} = m - \Sigma_{n,p}(T,n,\delta) \: .
\end{equation}
%and
%\begin{equation}
% m_{p}^{\ast} = m_{p} - S(T,n,-\delta)/c^{2} \: ,
%\end{equation}
%respectively.
Approximations for the functions $\Sigma_{n,p}^{0}(T,n,\delta)$ and 
$\Sigma_{n,p}(T,n,\delta)$ are given in the Appendix.
These functions reproduce the empirical values for the saturation density
$n_{\rm sat}\approx 0.15$~fm$^{-3}$ 
and the binding energy per nucleon
$B/A\approx -16$~MeV, see Subsection \ref{subsec:mp}. The
effective mass is somewhat smaller than the empirical value 
$m^{\ast} \approx m(1-0.17~n/n_{\rm sat})$ for $n<0.2$ fm$^{-3}$.

\subsection{Medium modification of cluster properties}
\label{subsec:medium2}

Recent progress of the description of clusters in low density nuclear
matter \cite{Ropke:2005,Ropke:2006,Sumiyoshi:2008qv,Ropke:2008qk}
enables us to evaluate the properties 
of deuterons, tritons, helions
and helium nuclei in a non-relativistic 
microscopic approach, taking the influence of the
medium into account. 

In addition to the $\delta$-like nucleon quasiparticle contribution, also the
contribution of the bound and scattering states
can be included in the single-nucleon spectral function by analyzing the
imaginary part of $\Sigma (1,z)$. Within a cluster decomposition,
$A$-nucleon $T$ matrices appear in a many-particle approach. These $T$
matrices describe the propagation of the $A$-nucleon cluster in nuclear
matter. In this way, bound states contribute to  $n_{\tau} =
n_{\tau}(T,\tilde{\mu}_{n}, \tilde{\mu}_{p})$, 
see \cite{Ropke:1983,Schmidt:1990}. 
Restricting the cluster decomposition only to the contribution of two-particle 
correlations, we obtain the so-called $T_{2}G$ approximation. 
In this approximation, the Beth-Uhlenbeck formula is obtained for the EoS, as 
shown in \cite{Ropke:1983,Schmidt:1990}. 
In the low-density limit, the propagation of the $A$-nucleon cluster is 
determined by the energy eigenvalues of the corresponding nucleus, and the 
simple EoS (\ref{EoS:p}) results describing the nuclear statistical 
equilibrium (NSE).

For nuclei imbedded in nuclear matter, an effective wave equation
can be derived \cite{Ropke:1983,Ropke:2008qk}.
The $A$-particle wave function $\psi_{A\nu K}(1\dots A)$
and the corresponding eigenvalues $E^{\rm qu}_{A, \nu}(K)$
follow from solving the in-medium Schr\"{o}dinger equation 
\begin{eqnarray}
\lefteqn{[E^{\rm qu}(1)+\dots + E^{\rm qu}(A) - E^{\rm qu}_{A, \nu}(K)]
 \psi_{A\nu K}(1\dots A)}
\nonumber \\ &&
+\sum_{1^{\prime} \dots A^{\prime}}\sum_{i<j}[1-\tilde{f}(i)- \tilde{f}(j)]
 V(ij,i^{\prime}j^{\prime})\prod_{k \neq
  i,j} \delta_{kk^{\prime}}\psi_{A \nu K}(1^{\prime}\dots A^{\prime})=0\,.
\label{waveA}
\end{eqnarray}
This equation contains the effects of the medium in the single-nucleon
quasiparticle shifts as well as in the Pauli blocking terms. The 
$A$-particle wave function and energy depend on the total momentum
$K$ relative to the medium.

The in-medium Fermi distribution function 
$\tilde{f}(1)=\left(\exp\left\{\beta 
\left[E^{\rm qu}(1)-\tilde \mu_{1}\right]\right\} +1 \right)^{-1}$ 
contains the non-relativistic
effective chemical potential $\tilde{\mu}_{1}$ which is determined
by the total proton or neutron densities
(i.e.\ including those bound in clusters) calculated in quasiparticle
approximation,
$n_{\tau}^{\rm tot} = 
\Omega^{-1} \sum_{1} \tilde f(1) \delta_{\tau_{1},\tau}$ for 
the particles inside the volume $\Omega$. It describes the
occupation of the phase space neglecting any correlations in the medium.
The solution of the in-medium Schr\"odinger equation (\ref{waveA}) can
be obtained in the low-density region by perturbation theory. In
particular, the quasiparticle energy of the $A$-nucleon cluster with
$Z$ protons in the ground state follows as
\begin{equation}
E^{\rm qu}_{A,\nu}(K) = 
E^{\rm qu}_{A,Z}(K)= E_{A,Z}^{(0)}+\frac{K^{2}}{2 A m}+
\Delta E_{A,Z}^{\rm SE}(K)+\Delta E_{A,Z}^{\rm Pauli}(K) 
 + \Delta E_{A,Z}^{\rm Coul}(K) + \dots 
\label{finalshift}
\end{equation}
with various contributions. Besides the cluster binding energy in the vacuum
$E_{A,Z}^{(0)}$
and the kinetic term, the self-energy shift 
$\Delta E_{A,Z}^{\rm SE}(K)$, the Pauli shift $\Delta E_{A,Z}^{\rm
  Pauli}(K)$
and the the Coulomb shift $\Delta E_{A,Z}^{\rm Coul}(K)$ enter. The latter
can be evaluated for dense matter in the Wigner-Seitz
approximation \cite{Kolomiets:1997zzb,Shlomo:2005,Ropke:1984}.
It is given by 
\begin{equation}
\Delta E_{A,Z}^{\rm Coul}(K)=\frac{Z^2}{A^{1/3}} 
\frac{3}{5} \frac{e^{2}}{r_{0}} 
\left[ \frac{3}{2} \left(\frac{2 n_p}{n_{\rm sat}}\right)^{\frac{1}{3}} 
- \frac{n_p}{n_{\rm sat}}\right]  
\end{equation}
with $r_{0} = 1.2$~fm. Since the values of $Z$ are small, this 
contribution is small as well and disregarded here 
%together with other small terms 
in the quasiparticle energy (\ref{finalshift}).

The self-energy contribution to the quasiparticle shift is determined by the
contribution of the single-nucleon shift
\begin{equation}
\label{delArigid}
\Delta E_{A,Z}^{\rm SE}(0)= (A-Z) \Delta E_{n}^{\rm SE}(0)+ Z \Delta
E_{p}^{\rm SE}(0) +\Delta E_{A,Z}^{\rm SE, eff.mass}\: .
\end{equation}
The contribution to the self-energy shift due to the change of the effective 
nucleon mass can be calculated from perturbation theory
using the unperturbed wave function of the clusters, 
see \cite{Sumiyoshi:2008qv}, so that
\begin{equation}
\label{dSelSEeff}
%\Delta E_{A,Z}^{\rm SE,
%  eff.mass}=\left(\frac{m}{m^{\ast}}-1\right)s_{A,Z}\: .
 \Delta E_{A,Z}^{\rm SE, eff.mass}
 = \left(1-\frac{m^{\ast}}{m}\right)s_{A,Z} \: .
\end{equation}
Values of $s_{A,Z}$
for $ \{A,Z\}=\{i\}=\{d,t,h,\alpha\} $ are given in Tab. \ref{tab:01}.
Inserting the medium-dependent quasiparticle energies in the
distribution functions (\ref{faz})
% \begin{equation}
%  f_{A,Z}[E^{\rm qu}_{A,\nu}(P)]=
%   \left\{\exp \left[ \left( E^{\rm
%       qu}_{A,\nu}(P)-\tilde{\mu}_{A,Z}\right)/T \right]- (-1)^{A}
%  \right\}^{-1} \:,
% \end{equation}
the first two contributions to the quasiparticle shift in 
(\ref{delArigid}) can
be included renormalizing the chemical potentials.

The most important effect in
the calculation of the abundances of light
elements comes from the Pauli blocking terms in Eq.\ (\ref{waveA}) in
connection with the interaction potential. This contribution is
restricted only to the bound states so that it may lead to the
dissolution of the nuclei if the density of nuclear matter increases.
The corresponding shift $\Delta E_{A,Z}^{\rm Pauli}(K)$ can be evaluated
in perturbation theory provided the interaction potential and the
ground state wave function are known.
After angular averaging where in the Fermi functions the mixed 
scalar product $\vec{k} \cdot \vec{K}$ between the total momentum 
 $\vec{K}$ 
and the remaining Jacobian coordinates $\vec{k}$ is neglected, 
the Pauli blocking shift can be approximated as
\begin{equation}
\label{delpauli0P}
\Delta E_{A,Z}^{\rm Pauli}(K) \approx \Delta E_{A,Z}^{\rm Pauli}(0) \,
\exp\left(-\frac{K^{2}}{2 A^{2} m T}\right) \: .
\end{equation}
Avoiding angular averaging, the full solution gives the result up to
the order $K^2$
\begin{equation}
\label{delpauli0P1}
\Delta E_{A,Z}^{\rm Pauli}(K) \approx \Delta E_{A,Z}^{\rm Pauli}(0) \,
\exp\left(-\frac{K^{2}}{g_{A,Z}}\right) 
\end{equation}
with the dispersion that can be calculated from
\begin{equation}
g_{i}(T,n,Y_{p})
= \frac{g_{i,1} + g_{i,2} T+h_{i,1} n}{1+h_{i,2} n} \: .
\end{equation}
The values for $g_{i,1}$ and $g_{i,2}$ 
can be calculated from perturbation theory
using the unperturbed cluster wave functions; the density corrections 
$h_{i,1}$ and $h_{i,2}$ 
are fitted to variational solutions of the in-medium wave equation
Eq.\ (\ref{waveA}) for given $T$, $n_p$, $n_n$ and $K$. 
Numerical values of the parameters in symmetric nuclear matter 
($Y_{p}=0.5$) are given in Tab \ref{tab:01}.

The shift of the binding energy of light clusters at zero total momentum
which is of first order in density  \cite{Ropke:2005,Ropke:2006} has been
calculated recently \cite{Ropke:2008qk}. 
The light clusters of the deuteron ($d={}^{2}$H), triton ($t={}^{3}$H), 
helion ($h={}^{3}$He) and the $\alpha$ particle (${}^{4}$He)
have been considered.
The interaction potential and the nucleonic
wave function of the few-nucleon system have been fitted to the binding
energies and the rms radii of the corresponding nuclei. 

With the neutron number $N_{i} = A_{i}-Z_{i}$, 
it can be written as
\begin{equation} \label{eq:lin_be_shift}
\Delta E_{A_{i},Z_{i}}^{\rm
  Pauli}(0;n_{p},n_{n},T) = 
 -\frac{2}{A_{i}}
 \left[Z_{i}n_{p}+N_{i}n_{n}\right] 
 \delta  E^{\rm
  Pauli}_{i}(T,n)~,
\end{equation}
where the temperature dependence and 
higher density corrections are contained in the functions
$ \delta  E^{\rm
  Pauli}_{i}(T,n)$. These functions have been obtained with different
approximations for the wave function.
In case of the deuteron, the Jastrow approach leads to a functional form
%\begin{equation}
% \delta B_{i}(T)  = 
% \frac{b_{i}}{(k_{B}T)^{3/2}}
%  \left[ \frac{1}{\sqrt{y_{i}}} -  \sqrt{\pi}c_{i}
% \exp \left( c_{i}^{2} y_{i} \right)
% {\rm erfc} \left(c_{i} \sqrt{y_{i}} \right)\right]
%\end{equation}
\begin{equation}
\label{eq:dP_d}
  \delta  E^{\rm
  Pauli}_{i}(T,n) = 
 \frac{a_{i,1}}{T^{3/2}}
  \left[ \frac{1}{\sqrt{y_{i}}} -  \sqrt{\pi}a_{i,3}
 \exp \left( a_{i,3}^{2} y_{i} \right)
 {\rm erfc} \left(a_{i,3} \sqrt{y_{i}} \right)\right] 
 \frac{1}{1+[b_{i,1}+b_{i,2}/T] n}
\end{equation}
with $y_{i} = 1+a_{i,2}/T$.
For the other clusters $i=t,h,\alpha$, the Gaussian approach is used
which gives the simple form
%\begin{equation}
%  \delta  E^{\rm Pauli}_{i}(T,n) = 
%  \frac{b_{i}}{\left( k_{B}T + a_{i} \right)^{3/2}} 
%\end{equation}
\begin{equation}
\label{eq:dP_tha}
\delta E_{i}^{\rm Pauli}(T,n)   = 
 \frac{a_{i,1}}{T^{3/2}}  \frac{1}{y_i^{3/2}} 
 \frac{1}{1+[b_{i,1}+b_{i,2}/T] n} \: .
\end{equation}
%with only two parameters $a_{i,1}$ and $a_{i,2}$ contained in $y_i$.
The parameters $a_{i,1}$, $a_{i,2}$ and $a_{i,3}$ 
are determined by low-density perturbation theory from the 
unperturbed cluster wave functions. The parameters $b_{i,1}$ and $b_{i,2}$ are 
density corrections and are fitted to the numerical solution of the 
in-medium wave equation Eq.\ (\ref{waveA}) 
for given $T,n_p,n_n,P=0$. Values are given in Tab \ref{tab:01}.

%\begin{table}[t]
%\caption{\label{tab:01}%
%Parameters for the cluster binding energy shifts.}
%\begin{tabular}{cdddddddddd}
%\hline \hline
% cluster $i$ & s_{i} 
% & a_{i,1}  & a_{i,2}  & a_{i,3} &
% b_{i,1} & b_{i,2} & g_{i,1}
%  & g_{i,2} & h_{i,1} & h_{i,2}  \\
%  & [\mbox{MeV}] & [\mbox{MeV}^{5/2}\mbox{fm}{}^{3}] & [\mbox{MeV}] & &
%  [\mbox{fm}^{3}] & [\mbox{MeV fm}^{3}] & [\mbox{fm}{}^{-2}] & 
%  [\mbox{MeV}^{-1}\mbox{fm}{}^{-2}] &
%  [\mbox{fm}] & [\mbox{fm}^{3}] \\
% \hline
% $d$    & 11.147  & 38386.41 & 22.52035 & 0.2223 & 1.048 & 285.7 &0.85 & 0.223 & 132 & 17.5\\
% $t$     &24.575 & 69516.2 & 7.49232 &-& 4.4142 &43.9 &3.2 & 0.45 & 37  & - \\
% $h$      &20.075& 58442.5 & 6.07718 &- & 4.4142 &43.9 & 2.638 & 0.434 & 43 & -\\
% $\alpha$ & 49.868& 164371 & 10.6701 &-& - & - & 8.2355&0.7718 & 50 & - \\
%\hline \hline
%\end{tabular}
%\end{table}

\begin{table}[t]
\caption{\label{tab:01}%
Parameters for the cluster binding energy shifts.}
\begin{tabular}{ccccccccccc}
\hline \hline
 cluster $i$ &$s_{i}$ 
 & $a_{i,1}$  & $a_{i,2}$  &$a_{i,3}$ &
 $b_{i,1}$ & $b_{i,2}$  & $ g_{i,1}$
  &  $ g_{i,2}$   &  $ h_{i,1}$  &  $ h_{i,2}$  \\
  & [MeV] & [MeV${}^{5/2}$fm${}^{3}$] & [MeV] & &
  [fm${}^{3}$] & [MeV fm${}^{3}$] & [fm${}^{-2}$] & [MeV${}^{-1}$fm${}^{-2}$] &
  [fm] & [fm${}^{3}$] \\
 \hline
 $d$  & 11.147  & 38386.4 & 22.5204 & 0.2223& 1.048 & 285.7 &0.85 & 0.223 & 132 & 17.5\\
 $t$  & 24.575 & 69516.2 & 7.49232 &-& 4.414 &43.90 &3.20 & 0.450 & 37  & - \\
 $h$  & 20.075& 58442.5 & 6.07718 &- & 4.414 &43.90 & 2.638 & 0.434 & 43 & -\\
 $\alpha$ & 49.868& 164371 & 10.6701 &-& - & - & 8.236&0.772 & 50 & - \\
\hline \hline
\end{tabular}
\end{table}

Now, the nucleon number densities 
(\ref{quasigas}) can be evaluated as in the non-interacting case, 
with the only difference that
the number densities of the particles are calculated with the 
quasiparticle energies.
In the light cluster-quasiparticle approximation, the total densities of 
neutrons
\begin{equation}
 n_{n}^{\rm tot} = n_{n} + \sum_{i=d,t,h,\alpha} N_{i} n_{i}
\end{equation}
and of protons
\begin{equation}
 n_{p}^{\rm tot} = n_{p} + \sum_{i=d,t,h,\alpha} Z_{i} n_{i}
\end{equation}
contain the densities of the free neutrons and protons 
$n_{n}$ and $n_{p}$,
respectively, and the contributions from the nucleons bound in the
clusters with densities $n_{i}$. 
The state of the system in chemical equilibrium
is completely determined by specifying the
total nucleon density $n=n_{n}^{\rm tot}+n_{p}^{\rm tot}$,
the asymmetry $\delta$ % = (n_{n}^{\rm tot}-n_{p}^{\rm tot})/n$ 
and the temperature $T$ as long as no $\beta$-equilibrium is considered.

This result is an improvement of the NSE and allows for the smooth
transition from the low-density limit up to the region of saturation
density. The bound state contributions to the EoS are fading with
increasing density because they move as resonances into 
the continuum of scattering
states. This improved NSE, however, does not contain the contribution of
scattering states explicitly.
For the treatment of continuum states in the two-nucleon case, as well as the
evaluation of the second virial coefficient, see 
\cite{Schmidt:1990,Horowitz:2005nd}.

The account of scattering states needs further consideration. 
Investigations on the two-particle level have been performed and extensively 
discussed \cite{Ropke:1983,Schmidt:1990,Horowitz:2005nd}.
We use the Levinson theorem to take the contribution of scattering 
states into account in the lowest-order
approximation. Each bound state
contribution to the density has to accompanied with a continuum
contribution that partly compensates the strength of the bound state
correlations.
As a consequence, the total proton and neutron densities are given by
\begin{eqnarray}
\label{quasigas2_p}
 n^{\rm tot}_{p}(T,\tilde{\mu}_p,\tilde{\mu}_n)&=& 
 \frac{1 }{\Omega} \sum_{A,\nu,K}^{\rm bound} Z 
\left[  f_{A,Z}[E^{\rm qu}_{A,\nu}(K;T,\tilde{\mu}_p,\tilde{\mu}_n)]
- f_{A,Z}[E^{\rm cont}_{A,\nu}(K;T,\tilde{\mu}_p,\tilde{\mu}_n)] \right]\,,
  \\  
\label{quasigas2_n}
 n^{\rm tot}_{n}(T,\tilde{\mu}_p,\tilde{\mu}_n)&=& 
 \frac{1}{\Omega} \sum_{A,\nu,K}^{\rm bound} (A-Z) 
\left[  f_{A,Z}[E^{\rm qu}_{A,\nu}(K;T,\tilde{\mu}_p,\tilde{\mu}_n)]
- f_{A,Z}[E^{\rm cont}_{A,\nu}(K;T,\tilde{\mu}_p,\tilde{\mu}_n)] \right]\,
\end{eqnarray}
with explicit bound and scattering terms.
 $E^{\rm cont}_{A,\nu}$ denotes the edge of the 
continuum states that is also determined by the single-nucleon 
self-energy shifts. 
These expressions guarantee a smooth behavior when the bound 
states merge with the continuum of scattering states.
The summation over $A$ includes also the contribution of free nucleons, $A=1$, 
considered as quasiparticles with the energy dispersion 
given by the RMF approach. 

The summation over $K$ and the subtraction of the
continuum contribution is extended only over 
that region of momentum space where 
bound states exist. The disappearance 
of the bound states is caused by the Pauli blocking term; the self-energy
contributions to the quasiparticle shifts act on bound as well 
as on scattering states.
Above the so-called Mott density, where the bound states at $K = 0$ disappear,
the momentum summation has to be extended only over that region
$K > K^{\rm Mott}_{A,\nu}(T,n,\delta)$ where the bound state energy 
is lower than the
continuum of scattering states.
The contribution of scattering states is necessary to obtain the second
virial coefficient according to the Beth-Uhlenbeck equation, see
\cite{Horowitz:2006pj,Schmidt:1990}. 
This leads also to corrections in comparison
with the NSE that accounts only for the bound state contributions,
neglecting all effects of scattering states. These corrections become 
important at increasing temperatures for weakly bound clusters. Thus,
the corrections which lead to the correct second virial coefficient are
of importance for the deuteron system, when the temperature is
comparable or large compared with the binding energy per nucleon. In the
calculations for the quantum statistical (QS) model
shown below, the contributions of these continuum
correlations have been taken into account.

Solving Eqs. (\ref{quasigas2_p}) and (\ref{quasigas2_n})
for given $T$, $n^{\rm tot}_{p}$ and 
$n^{\rm tot}_{n}$ we find the
chemical potentials $\mu_p$ and $\mu_n$. After integration, 
see Eq. (\ref{eq:freV}), the free
energy is obtained, and all the other thermodynamic 
functions are derived from this quantity
without any contradictions. Results are given below.

We do not consider the formation of heavy clusters here. 
This limits the parameter range $n_{n}^{\rm tot}$, $n_{p}^{\rm  tot}$,
$T$ in the phase diagram to that area where the 
abundances of heavier clusters are small. 
For a more general approach to the EoS which
takes also the contribution of heavier cluster into account, see
\cite{Ropke:1984}.
Future work will include the contribution of the heavier clusters. 

Further approximations refer to the linear dependence on density of the shifts 
of binding energies, calculated in perturbation theory. 
A better treatment will improve these shifts, but it can be shown that the 
changes are small. 
The approximation of the uncorrelated medium can be improved 
considering the cluster mean-field approximation 
\cite{Ropke:1980,Ropke:1983,Ropke:2008qk}. 
Furthermore, the formation of quantum condensates will give additional
contributions to the EoS. However, in the region considered here the formation
of quantum condensates does not appear. This is in
contrast to a recent work employing a quasiparticle gas model
\cite{Heckel:2009br} where Bose-Einstein condensation of deuterons is observed
because the Pauli shift of the deuteron binding energy at high densities is
not considered.

\section{Generalized relativistic mean-field model with light clusters}
\label{sec:RMF}

A main ingredient to construct the low-density EoS is the proper determination 
of the nucleonic quasiparticle energies that enter the single-nucleon 
distribution functions, Eqs.\ (\ref{nqu},\ref{quasigas})
but also the cluster 
energies via the in-medium Schr\"odinger equation (\ref{waveA}). 
Recently, realistic values for the nucleon quasiparticle shifts were obtained 
from sophisticated calculations within Hamiltonian approaches, such 
as Dirac-Brueckner Hartree-Fock calculations 
\cite{Fuchs:2005yn,vanDalen:2005ns}. 
RMF approaches proved to be very successful to interpret properties near 
saturation density, see, e.g., Refs.\
\cite{Typel:2005ba,Klahn:2006ir}. 
We extract the single-nucleon quasiparticle 
shifts from the results of the RMF model with density-dependent
couplings and use them in our QS approach. 
In the following we show how this RMF model 
can be extended to include light clusters, which are considered 
as quasiparticles modified by medium effects as obtained in the QS
approach. A comparison of the generalized RMF model with the QS
model will show distinct differences in the thermodynamical
properties that are related to the employed approximations.

In a conventional relativistic mean-field description 
\cite{Serot:1984ey}
of homogeneous and isotropic nuclear matter, 
nucleons interact by the exchange of mesons where usually isoscalar $\omega$
and $\sigma$ and isovector $\rho$ mesons are included. Neutrons and protons
are described by Dirac spinors $\psi_{i}$ ($i=n,p$). The mesons 
are represented by Lorentz vector fields $\omega_{\mu}$ and
$\vec{\rho}_{\mu}$ and Lorentz scalar fields $\sigma$. The
electromagnetic interaction is not considered in nuclear matter. 
A possible isovector,
Lorentz scalar $\delta$ meson is not included in the present model.
The mesons couple minimally to the nucleons. In our approach, 
non-linear meson self-interactions are not introduced, 
but the couplings are
assumed to be functionals of the nucleon operator-valued currents 
in order to simulate
a medium dependence of the interaction.

In the generalized RMF model with light clusters, the ground states
of the deuteron ($d={}^{2}$H), the triton ($t={}^{3}$H), the
helion ($h={}^{3}$He) and the $\alpha$ particle (${}^{4}$He) are 
introduced as additional degrees of freedom with the corresponding
spin $0$ field $\phi_{\alpha}$, spin $1$ field $\phi_{d}^{\nu}$ and
spin $1/2$ fields $\psi_{i}$ ($i=t,h$). The clusters are treated as
point-like particles and their internal structure is not taken into
account. The influence of the medium on
the cluster properties is described by density and temperature
dependent shifts of the binding energies as introduced in the previous
section. The Pauli shifts, cf.\ Eq.\ (\ref{eq:lin_be_shift}), 
are taken from the
nonrelativistic calculation neglecting the dependence
on the c.m.\ momentum $K$ of the cluster, while this is taken into
account in the QS approach.
On the other hand, the self-energy shift is treated
self-consistently in the RMF description in contrast to the QS
approach where it enters in parametrized form from an independent model,
namely the RMF model decribed in this section.

\subsection{Lagrangian density and field equations}

In the present approach, the model Lagrangian has the form
\begin{eqnarray}
\label{eq:Lag}
 \mathcal{L} & = & 
\sum_{i=n,p,t,h}
\bar{\psi}_{i} \left( \gamma_{\mu} iD_{i}^{\mu} - M_{i} \right) \psi_{i}
%\\ \nonumber & & 
+ \frac{1}{2} 
 \left( iD_{\alpha}^{\mu} \varphi_{\alpha} \right)^{\ast}
 \left( iD_{\alpha\mu} \varphi_{\alpha} \right)
 - \frac{1}{2} \varphi_{\alpha}^{\ast} M_{\alpha}^{2} \varphi_{\alpha}
 \\ \nonumber & & 
 + \frac{1}{4} \left( iD_{d}^{\mu} \varphi_{d}^{\nu}
  - iD_{d}^{\nu} \varphi_{d}^{\mu} \right)^{\ast}
  \left( iD_{d\mu} \varphi_{d\nu}
  - iD_{d\nu} \varphi_{d\mu} \right)
 - \frac{1}{2} \varphi_{d}^{\mu\ast} M_{d}^{2} \varphi_{d\mu}
 \\ \nonumber & & + \frac{1}{2} \left( 
 \partial^{\mu} \sigma \partial_{\mu} \sigma - m_{\sigma}^{2} \sigma^{2}
 - \frac{1}{2} G^{\mu\nu} G_{\mu\nu} + m_{\omega}^{2} \omega^{\mu} \omega_{\mu}
 - \frac{1}{2} \vec{H}^{\mu\nu} \cdot \vec{H}_{\mu\nu} 
 + m_{\rho}^{2} \vec{\rho}^{\mu} \cdot \vec{\rho}_{\mu}
% - \frac{1}{2} F^{\mu\nu} F_{\mu\nu}
 \right)
\end{eqnarray}
with the field tensors
\begin{equation}
% F_{\mu\nu} = \partial_{\mu} A_{\nu} -  \partial_{\nu} A_{\mu}
% \qquad
 G_{\mu\nu} = \partial_{\mu} \omega_{\nu} -  \partial_{\nu} \omega_{\mu}
 \qquad
 \vec{H}_{\mu\nu} = \partial_{\mu} \vec{\rho}_{\nu} -  
 \partial_{\nu} \vec{\rho}_{\mu}  
% \qquad
% \Phi_{d\mu\nu} = \partial_{\mu} \varphi_{d\nu} 
% - \partial_{\nu} \varphi_{d\mu}
\end{equation}
of the Lorentz vector fields.
Vectors in isospin space carry an arrow. Nucleons form an isospin
doublet with $\tau_{3}\psi_{n} = \psi_{n}$ and
$\tau_{3}\psi_{p} = -\psi_{p}$. Similarly, for the triton and
helion one has $\tau_{3}\psi_{t} = \psi_{t}$ and
$\tau_{3}\psi_{h} = -\psi_{h}$, respectively.
Deuterons and $\alpha$-particles are treated as isospin
singlets. 

The covariant derivative 
\begin{equation}
 i D_{i}^{\mu} = i \partial^{\mu}
 - \Gamma_{\omega} A_{i} \omega^{\mu} 
 - \Gamma_{\rho} |N_{i}-Z_{i}| \vec{\tau} \cdot \vec{\rho}^{\mu} 
\end{equation}
for a particle $i$ contains the interaction with the Lorentz
vector mesons with a strength that is determined by the 
density-dependent couplings
$\Gamma_{\omega}$, $\Gamma_{\rho}$
and the mass ($A_{i}$), neutron
($N_{i}$) and proton ($Z_{i}$)
number of a particle $i$.
The scalar $\sigma$ meson with coupling strength $\Gamma_{\sigma}$
appears in the effective mass 
\begin{equation}
 M_{i} = m_{i} - \Gamma_{\sigma} A_{i} \sigma - \Delta B_{i}
\end{equation}
of a particle $i$ with vacuum rest mass $m_{i}$.
The vacuum rest mass
of a cluster $i=d,t,h,\alpha$ 
is given by
\begin{equation}
 \label{eq:mass_cl}
 m_{i} =  Z_{i}m_{p} + N_{i}m_{n} - B_{i}^{0} \: ,
\end{equation}
which defines the vacuum binding energies $B_{i}^{0}>0$.
The medium dependent Pauli shift $\Delta B_{i}$
appears only for clusters.
The couplings $\Gamma_{m}=\Gamma_{m}(\varrho)$ ($m=\omega,\sigma,\rho$) are 
functionals of the Lorentz scalar density
\begin{equation}
 \varrho = \sqrt{J^{\mu}J_{\mu}} 
\end{equation}
that contains the free nucleon current
\begin{equation}
 J^{\mu} =  j^{\mu}_{p} + j^{\mu}_{n}
\end{equation}
with $j^{\mu}_{i} = \bar{\psi}_{i} \gamma^{\mu} \psi_{i}$.
The Pauli shifts of the binding energies were derived in the
previous section as a function
$\Delta B_{i}(n_{p}^{\rm tot},n_{n}^{\rm tot},T)$ 
depending on the total proton and
neutron densities $n_{p}^{\rm tot}$ and $n_{n}^{\rm tot}$
and the temperature $T$.
In principle, the densities have to be replaced by the
corresponding quantities expressed in terms of the field operators 
of the nucleons and clusters. In the case of the fermions, this
poses no problem since the currents of the triton and helion
have the same form as the currents of the nucleons. However, for the
bosons, the definition of the conserved currents 
contains the meson fields and the Pauli shifts. 
A dependence of the Pauli shifts on these currents
leads to additional rearrangement contributions
that modify the self-energies and requires a redefinition
of the conserved currents. In order to avoid these
complications, the dependence of the Pauli shifts on the densities
is replaced by a dependence on the vector meson fields.
In this way the Pauli shifts are treated in an equivalent way 
as the usual single-particle shifts. In consequence, we replace
the dependence on the total densities
\begin{eqnarray}
 n_{n}^{\rm tot} & \to & n_{n}^{\rm ps} = \frac{1}{2} 
 \left[\varrho_{\omega} + \varrho_{\rho}\right]
 \\
 n_{p}^{\rm tot} & \to & n_{p}^{\rm ps} = \frac{1}{2} 
 \left[\varrho_{\omega} - \varrho_{\rho}\right]
\end{eqnarray}
by pseudo-densities $n_{n}^{\rm ps}$ and $n_{p}^{\rm ps}$
with the quantities
\begin{equation}
 \varrho_{\omega} = \lambda_{\omega} \sqrt{\omega^{\mu}\omega_{\mu}}
\end{equation}
and
\begin{equation}
 \varrho_{\rho} = \lambda_{\rho}
 \sqrt{\vec{\rho}^{\mu}\cdot\vec{\rho}_{\mu}}
 \: .
\end{equation}
The coefficients $\lambda_{\omega}=m_{\omega}^{2}/\Gamma_{\omega}(0)$ 
and $\lambda_{\rho}=m_{\rho}^{2}/\Gamma_{\rho}(0)$ are
defined such that the correct low-density limit is obtained.

The field equations for mesons, nucleons and clusters are derived from
the Lagrangian density in the usual way. They are solved 
self-consistently in the mean-field
approximation where the meson fields are treated as classical fields
and sea-states of the fermions are not considered.
The couplings $\Gamma_{m}$ ($m=\omega,\sigma,\rho$)
become simple functions of 
$\varrho = \sqrt{\langle J^{\mu} \rangle \langle J_{\mu} \rangle}$
where the brackets 
$\langle \cdot \rangle$ indicate
the summation over all occupied states of the system.
The field equations simplify considerably 
due to the symmetries of homogeneous and isotropic nuclear matter at rest. 
The scalar meson field is directly given by
\begin{equation}
 \sigma = \frac{\Gamma_{\sigma}}{m_{\sigma}^{2}} n_{\sigma}
\end{equation}
with the source density
\begin{equation}
 n_{\sigma} = \sum_{i=n,p,d,t,h,\alpha} A_{i} n_{i}^{s}
\end{equation}
that is a sum of the scalar densities 
$n_{i}^{s}=\langle \bar{\psi}_{i} \psi_{i} \rangle$
of the fermions $i=n,p,t,h$ and the scalar densities
$n_{d}^{s} = \langle \varphi_{d}^{\mu} M_{d}\varphi_{d\mu} \rangle$
of the deuteron and 
$n_{\alpha}^{s} = \langle \varphi_{\alpha} M_{\alpha}\varphi_{\alpha} \rangle$
of the $\alpha$ particle.
The non-vanishing components of the vector meson fields are
\begin{eqnarray}
 \label{eq:omega}
 \omega_{0} & = &  \frac{\Gamma_{\omega}}{m_{\omega}^{2}}
 n_{\omega} - \sum_{i=d,t,h,\alpha} \frac{\lambda_{\omega}}{2m_{\omega}^{2}} 
 \left( \frac{\partial \Delta B_{i}}{\partial n_{n}^{\rm ps}}
 + \frac{\partial \Delta B_{i}}{\partial n_{p}^{\rm ps}} \right) n_{i}^{s}
 \\ 
 \label{eq:rho}
 \left(\vec{\rho}_{0}\right)_{3}
 & = & \frac{\Gamma_{\rho}}{m_{\rho}^{2}}
 n_{\rho} - \sum_{i=d,t,h,\alpha} \frac{\lambda_{\rho}}{2m_{\rho}^{2}} 
 \left( \frac{\partial \Delta B_{i}}{\partial n_{n}^{\rm ps}}
 - \frac{\partial \Delta B_{i}}{\partial n_{p}^{\rm ps}} \right) n_{i}^{s}
\end{eqnarray}
with two different source contributions. The regular source densities
\begin{eqnarray}
 n_{\omega} & = & \sum_{i=n,p,d,t,h,\alpha} A_{i} n_{i}
 \\
 n_{\rho} & = & \sum_{i=n,p,d,t,h,\alpha} \left(N_{i}-Z_{i}\right) n_{i}
\end{eqnarray}
depend on the vector densities 
$n_{i}=\langle \bar{\psi}_{i} \gamma_{0} \psi_{i} \rangle$ of the
fermions $i=n,p,t,h$, the vector density of the
deuteron 
\begin{equation}
 n_{d} = \frac{1}{2} \langle 
 \left( iD_{d0} \varphi_{d\mu} - iD_{\mu}\varphi_{d0}\right)^{\ast}
 \varphi_{d}^{\mu} + \varphi_{d}^{\mu \ast}
 \left( iD_{d0} \varphi_{d\mu} - i D_{d\mu} \varphi_{d0}\right)
 \rangle
\end{equation}
and of the $\alpha$ particle
\begin{equation}
 n_{\alpha} = \frac{1}{2} \langle 
 \left( iD_{\alpha 0} \varphi_{\alpha}\right)^{\ast}
 \varphi_{\alpha} + \varphi_{\alpha}^{\ast}
 \left( iD_{\alpha 0} \varphi_{\alpha}\right)
 \rangle \: .
\end{equation}
The second contribution in eqs.\ (\ref{eq:omega}) and (\ref{eq:rho})
with the derivatives of the binding energy shifts 
is proportional to the scalar densities of the clusters.

The Dirac equation for the spin $1/2$ particles ($i=n,p,t,h$) 
assumes the form
\begin{equation}
  \left[ \gamma^{\mu} \left( i \partial_{\mu} -
      \Sigma_{i\mu} \right)
 - \left( m_{i} - \Sigma_{i} \right) 
 \right] \psi_{i} = 0 
\end{equation}
with scalar and vector self-energies $\Sigma_{i}$ and $\Sigma_{i\mu}$,
respectively. The field equations for the $\alpha$-particle and the
deuteron are the Klein-Gordon equation
\begin{equation}
 - \left[ \left( i\partial_{\mu}- \Sigma_{\alpha\mu} \right)
 \left( i\partial^{\mu}- \Sigma_{\alpha\mu} \right)
  + \left( m_{\alpha}-\Sigma_{\alpha}\right)^{2} \right]
 \varphi_{\alpha} = 0
\end{equation}
and the Proca equation
\begin{equation}
 - \left( i\partial_{\mu} - \Sigma_{d\mu} \right)
 \left[  \left( i\partial^{\mu} 
 - \Sigma_{d}^{\mu}\right) \varphi_{d}^{\nu} 
 - \left( i \partial^{\nu}
 - \Sigma_{d}^{\nu} \right) \varphi_{d}^{\mu} \right]
 + \left( m_{d}-\Sigma_{d} \right)^{2} \varphi_{d}^{\nu} = 0 \: ,
\end{equation}
respectively. %, with only scalar self-energies.
%For nucleons ($i=n,p$) the self-energies are given by
The scalar self-energies are given by
\begin{equation}
  \Sigma_{i} = \Gamma_{\sigma} A_{i} \sigma  + \Delta B_{i}
\end{equation}
where the binding energy shift $\Delta B_{i}$ appears only for clusters.
The non-vanishing component of the vector self-energy is the
zero-component 
\begin{equation}
  \Sigma_{i0} = 
   \Gamma_{\omega} A_{i} \omega_{0} 
 + \Gamma_{\rho} \left( N_{i} - Z_{i} \right) \left(\vec{\rho}_{0}\right)_{3} 
 + \Sigma_{i0}^{R} 
\end{equation}
with the `rearrangement' contribution
\begin{eqnarray}
 \Sigma_{i0}^{R} & = & 
 \Gamma_{\omega}^{\prime} \omega_{0} n_{\omega}
 + \Gamma_{\rho}^{\prime} \left(\vec{\rho}_{0} \right)_{3} n_{\rho}
  - \Gamma_{\sigma}^{\prime} \sigma n_{\sigma}
% - \sum_{j=d,t,h,\alpha} n^{s}_{j}  \frac{\partial
%   (\Delta B_{j})}{\partial n_{i}} 
\end{eqnarray}
that appears only for nucleons.
It contains contributions with derivatives 
%\begin{equation}
$ \Gamma_{m}^{\prime} = d\Gamma_{m}/d\varrho$
%\end{equation}
of the meson-nucleon couplings.

Since the self-energies are momentum-independent in homogeneous 
and isotropic nuclear 
matter, the field equations of the nucleons and of the clusters are easily 
solved. 
The solutions are plane waves with shifted masses and energies
as compared to the vacuum solution, i.e.\ nucleons and clusters
can be considered as quasiparticles. 
At finite temperatures $T$ the relevant vector and scalar densities
are easily calculated with these solutions by integrating
over all momenta with the correct distribution functions.
Thus, the vector and scalar densities of the fermions ($i = n,p,t,h$)
are given by
\begin{eqnarray}
\label{eq:n_i_f}
 n_{i} & = &  
 g_{i} \int\frac{d^{3}k}{(2\pi)^{3}} \: 
 \left[  f_{i}^{+}(k) - f_{i}^{-}(k) \right]
 \\
\label{eq:n_i_f_s}
 n_{i}^{s} & = &  
 g_{i} \int \frac{d^{3}k}{(2\pi)^{3}} \: 
 \frac{m_{i}-\Sigma_{i}}{e_{i}(k)}
 \left[  f_{i}^{+}(k) + f_{i}^{-}(k) \right]
\end{eqnarray}
with degeneracy factor $g_{i}=2$ and the energy
\begin{equation}
 e_{i}(k) = \sqrt{k^{2}+\left(m_{i}-\Sigma_{i}\right)^{2}} \: .
\end{equation}
The Fermi-Dirac distribution for the particle ($\eta = 1$)
and antiparticle ($\eta = -1$) contributions is defined by
\begin{equation}
 f_{i}^{\eta}(k) = 
 \left\{\exp\left[ \eta \beta \left( E_{i}^{\eta} - \mu_{i}
 \right) \right] + 1 \right\}^{-1}
\end{equation}
where $\beta = 1/T$ and
%\begin{equation}
$ E_{i}^{\eta}(k) = 
 \Sigma_{i0} + \eta e_{i}(k)$ %\sqrt{k^{2}+\left(m_{i}-\Sigma_{i}\right)^{2}}
%\end{equation}
is the quasiparticle energy.
The (relativistic) chemical potential of a particle $i$
is denoted by $\mu_{i}$.
The densities of the bosons ($i = d, \alpha$) are obtained from
\begin{eqnarray}
\label{eq:n_i_b}
 n_{i} & = &  g_{i} \int \frac{d^{3}k}{(2\pi)^{3}} \: 
 b_{i}(k) + \tilde{n}_{i}
 \\
\label{eq:n_i_b_s}
 n_{i}^{s} & = &  g_{i} \int \frac{d^{3}k}{(2\pi)^{3}} \: 
  \frac{m_{i}-\Sigma_{i}}{e_{i}(k)} \:
 b_{i}(k) + \tilde{n}_{i}^{s}
\end{eqnarray}
with the Bose-Einstein distribution
\begin{equation}
 b_{i}(k) = \left\{\exp\left[ \beta \left( E_{i}^{+} - \mu_{i}
 \right) \right] - 1 \right\}^{-1}
\end{equation}
and degeneracy factors $g_{d} = 3$ and $g_{\alpha} = 1$, respectively.
A possible contribution to the densities from particles that are
condensed in the ground state is denoted by $\tilde{n}_{i}$ and
$\tilde{n}_{i}^{s}$. In homogeneous and isotropic matter these two are actually
identical. For a system of nucleons and clusters in chemical equilibrium, the
(relativistic) chemical potential of a cluster $i$ is determined by
\begin{equation}
 \mu_{i} = N_{i} \mu_{n} +  Z_{i} \mu_{p} \: .
\end{equation}
Thus, there are only two independent chemical potentials.

For given total baryon number density $n$, asymmetry $\delta$ and
temperature $T$, the coupled field equations of the generalized RMF
model are solved selfconsistently. 
This procedure yields the chemical potentials of
neutrons and protons that determine the densities of all particles.
Finally, all thermodynamical quantities, that are specified in the following
subsection, can be calculated.

\subsection{Thermodynamical quantities}

The energy density $\varepsilon$ and the pressure $p$ are derived from the
energy-momentum tensor $T^{\mu\nu}$ with the results
\begin{eqnarray}
 \varepsilon = \langle T^{00} \rangle & = &  
   \sum_{i = n,p,t,h}  g_{i} \int \frac{d^{3}k}{(2\pi)^{3}} \:
 \sum_{\eta}  f_{i}^{\eta} e_{i}(k)
%   \sqrt{k^{2}+\left(m_{i}-\Sigma_{i}\right)^{2}}
% \\ \nonumber & & 
 + \sum_{i=d,\alpha} \left[ g_{i} \int \frac{d^{3}k}{(2\pi)^{3}} \:
 b_{i} e_{i}(k) %\sqrt{k^{2}+\left(m_{i}-\Sigma_{i}\right)^{2}}
 + \tilde{n}_{i} \left( m_{i} - \Sigma_{i} \right) \right]
 \\ \nonumber & & 
 + \Gamma_{\omega} \omega_{0} n_{\omega}
 + \Gamma_{\rho} \rho_{0} n_{\rho}
 + \frac{1}{2}  \left[ 
 m_{\sigma}^{2} \sigma^{2}
 - m_{\omega}^{2}  \omega_{0}^{2}
 - m_{\rho}^{2} \rho_{0}^{2}  \right]
\end{eqnarray}
and
\begin{eqnarray}
 \label{eq:pres}
 p = \frac{1}{3} \sum_{m=1}^{3} \langle T^{mm} \rangle & = & 
\frac{1}{3} \sum_{i=n,p,t,h} g_{i} \int \frac{d^{3}k}{(2\pi)^{3}} \:
 \sum_{\eta} f_{i}^{\eta} 
 \frac{k^{2}}{e_{i}(k)} %\sqrt{k^{2}+\left(m_{i}-\Sigma_{i}\right)^{2}}}
+ \frac{1}{3} \sum_{i=d,\alpha} g_{i}  \int \frac{d^{3}k}{(2\pi)^{3}} \:
 b_{i} \frac{k^{2}}{e_{i}(k)} %\sqrt{k^{2}+\left(m_{i}-\Sigma_{i}\right)^{2}}}
 \\ \nonumber & & 
 + \left( n_{n} + n_{p} \right) \left[
 \Gamma_{\omega}^{\prime} \omega_{0} n_{\omega} %\left( n_{n} + n_{p} \right)
 +  \Gamma_{\rho}^{\prime}  \rho_{0} n_{\rho} % \left( n_{n} - n_{p} \right)
 - \Gamma_{\sigma}^{\prime} \sigma n_{\sigma} 
% \left( n_{n}^{s} + n_{p}^{s} \right)
 \right]
% \\ \nonumber & & 
% - \sum_{i=p,n} n_{i}
% \sum_{j=d,t,h,\alpha} 
% \frac{\partial (\Delta B_{j})}{\partial n_{i}} n_{j}^{s}
 - \frac{1}{2}  \left[   m_{\sigma}^{2} \sigma^{2}
 - m_{\omega}^{2}  \omega_{0}^{2} - m_{\rho}^{2} \rho_{0}^{2}
 \right]
\end{eqnarray}
with $\rho_{0} = \left( \vec{\rho_{0}}\right)_{3}$.
The condensed bosons do not contribute to the pressure but to the
energy density. 
The entropy density $s$ can be extracted from the
grand-canonical potential density $\omega(T,\mu_{n},\mu_{p}) = -p$ as 
$s =  - \left. \left( \partial \omega/\partial T 
\right)\right|_{\mu_{n},\mu_{p}}$.
However, it is more practical to use the thermodynamic relation
\begin{equation}
 \varepsilon =  Ts - p 
 + \sum_{i=n,p,d,t,h,\alpha} \mu_{i} n_{i} 
\end{equation}
corresponding to the Hugenholtz-van-Hove theorem.
After partial integration the standard result
\begin{eqnarray}
 s %=  - \left( \frac{\partial \omega}{\partial T}
 %\right)_{\mu_{i}} %,\xi_{j},\vec{\nabla}\xi_{j}}
 & = & 
- \sum_{i=n,p,t,h} g_{i} \int \frac{d^{3}k}{(2\pi)^{3}} \:
 \sum_{\eta} \left[ f_{i}^{\eta} \ln f_{i}^{\eta}
 + \left( 1- f_{i}^{\eta} \right) \ln \left( 1- f_{i}^{\eta} \right)
 \right] 
 \\ \nonumber & & 
 - \sum_{i=d,\alpha} g_{i} \int \frac{d^{3}k}{(2\pi)^{3}} \:
 \left[  b_{i} \ln b_{i}
 - \left( 1+ b_{i} \right) \ln \left( 1+ b_{i} \right) 
 \right] 
\end{eqnarray}
is obtained.
The thermodynamical pressure
\begin{equation}
 p = n^{2} \left. \frac{\partial}{\partial n} 
 \left( \frac{f}{n} \right) \right|_{T, \delta}
\end{equation}
calculated from the free energy density  $f = \varepsilon - Ts$
is identical to the pressure (\ref{eq:pres}) in the fieldtheoretical approach.
Since the energy density in the RMF model contains
the contribution of the rest mass of the particles, it is convenient
to define the internal energy per nucleon as
\begin{equation}
 \label{eq:EA}
 E_{A}(n,\delta,T) = \frac{1}{n} \left[ \varepsilon(n,\delta,T)
 - n_{n}^{\rm tot} m_{n} - n_{p}^{\rm tot} m_{p} \right] 
\end{equation}
and correspondingly the
free energy per nucleon 
\begin{equation}
 \label{eq:FA}
 F_{A}(n,\delta,T)=E_{A}-T\frac{s}{n} \: ,
\end{equation}
where the rest mass has been subtracted.
We emphasize that both the generalized RMF
model and the QS approach are thermodynamically consistent.

\subsection{Dissolution of clusters}

\begin{figure}
\epsfig{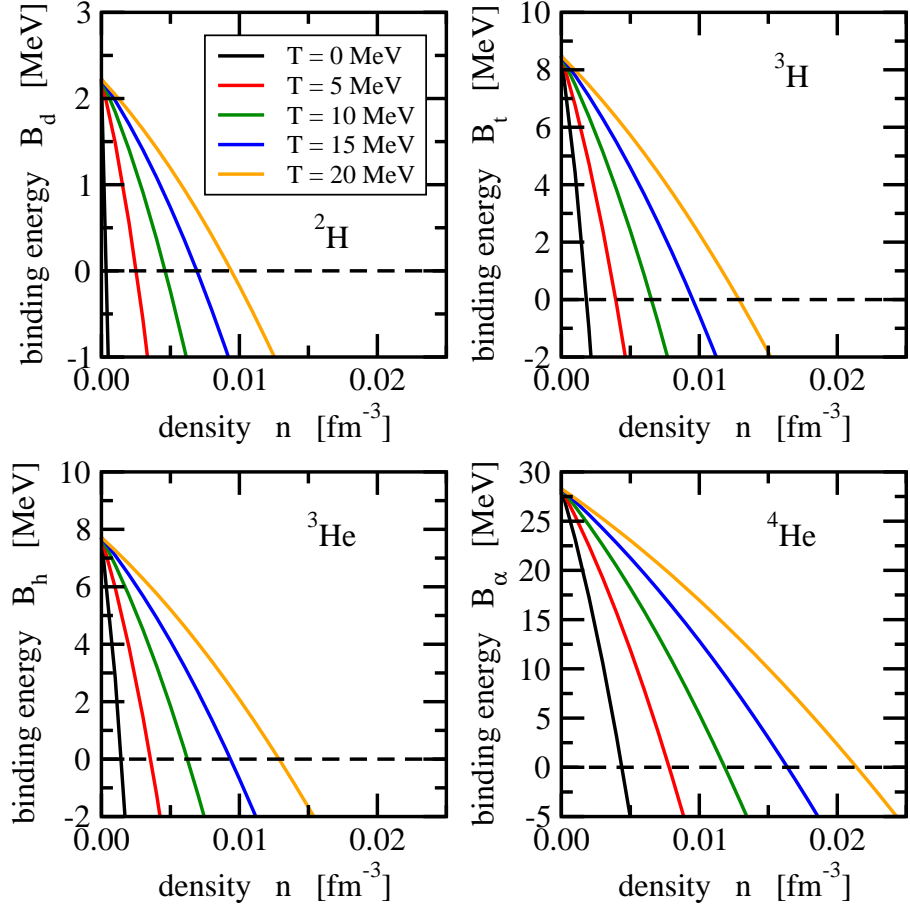}
\caption{\label{fig:00}%
(Color online)
Change of the binding energy $B_{i}=B_{i}^{0}+\Delta B_{i}$ 
of the clusters $i=d,t,h,\alpha$ at rest
in symmetric nuclear matter
due to the binding energy shift $\Delta B_{i}$ 
as used in the generalized
RMF model as a function of the total nucleon density
$n=n_{n}^{\rm tot}+n_{p}^{\rm tot}$ 
of the medium for various temperatures $T$.}
\end{figure}

%old version
%\begin{table}[t]
%\caption{\label{tab:01}%
%Parameters for the temperature dependence of the cluster binding
%energy shifts.}
%\begin{tabular}{cccc}
%\hline \hline
% cluster $i$ & $a_{i}$ [MeV] & $b_{i}$ [MeV${}^{5/2}$fm${}^{3}$] & $c_{i}$ \\
% \hline
% $d$      & 22.52035 & 38386.41 & 0.2223021\\
% $t$      & 7.492779 & 69521.59 \\
% $h$      & 6.077291 & 58451.64 \\
% $\alpha$ & 10.67040 & 164378.4 \\
%\hline \hline
%\end{tabular}
%\end{table}

In the generalized RMF approach, the quasiparticle energy shift of
the clusters contains the self-energy shift due to the mean fields
of the mesons. The effective-mass shift is already included in the 
relativistic approach and the Coulomb shift is neglected.
For the binding energy shift $\Delta B_{i}$
only the effect of the Pauli blocking is considered. 
The dependence of the Pauli shift on the cluster momentum relative to
the medium is neglected
since the introduction of a momentum dependence in the RMF Lagrangian
is non-trivial \cite{Typel:2005ba}.
With increasing density of the medium, the linear approximation
of the Pauli shift
(\ref{eq:lin_be_shift}) in the densities is not sufficient and
higher-order terms have to be considered. In the present RMF calculation,
an empirical quadratic form 
\begin{equation}
\label{eq:DBq}
 \Delta B_{i}(n_{p}^{\rm tot},n_{n}^{\rm tot},T) = - \tilde{n}_{i} 
 \left[ 1 + \frac{\tilde{n}_{i}}{2\tilde{n}_{i}^{0}(T)} \right]
 \delta B_{i} (T)
\end{equation}
is used, where the abbreviation
\begin{equation}
 \tilde{n}_{i} = \frac{2}{A_{i}}
 \left[Z_{i}n_{p}^{\rm tot}+N_{i}n_{n}^{\rm tot}\right]
\end{equation}
and the density scale 
\begin{equation}
 \tilde{n}_{i}^{0}(T) = \frac{B_{i}^{0}}{\delta B_{i}(T)}
\end{equation}
for the dissolution of the cluster $i$
with the vacuum binding energy $B_{i}^{0}$ are introduced.
The quantity $\delta B_{i}(T)$ is given by
$\delta E_{i}^{\rm Pauli}(T,0)$ in eq.\ (\ref{eq:dP_d}) for deuterons and
in eq.\ (\ref{eq:dP_tha}) for the other clusters.
In the limit $T \to 0$ the shifts and their derivatives remain finite
for all clusters.

The total binding energy of a cluster $i$ is the sum 
$B_{i}=B_{i}^{0}+\Delta B_{i}$ of the experimental binding energy $B_{i}^{0}$ 
in the vacuum  \cite{Audi:2002rp} and the binding energy shift
$\Delta B_{i}$ that in general depends on the c.m.\
momentum $K$ (see \ref{subsec:medium2}).
The dependence of the cluster binding energies on the
total nucleon density $n = n_{n}^{\rm tot}+n_{p}^{\rm tot}$ of the medium is
depicted for symmetric nuclear matter
in Fig.\ \ref{fig:00} for various temperatures $T$
and clusters at rest.
% It does not include the
%self-energy shift that affects both free and bound nucleons.}
For $n \to 0$ the experimental binding energy is recovered.
The density where a cluster becomes unbound, i.e.\ $B_{i}=0$, 
increases with increasing temperature. This behavior
is expected since the Pauli blocking of states is less effective
at higher temperatures. Note that both free nucleons and nucleons
bound in clusters are relevant for the Pauli principle
occupying phase space in momentum representation.
In principle, this is not described by a Fermi distribution but 
by the bound state wave functions. A more exact theory taking this 
into account is given by the cluster mean-field approximation 
\cite{Ropke:2008qk}, which, however, is very complex and has not 
been solved in general so far. 
Here we use the approximation of an effective Fermi distribution with an
effective chemical potential which includes both free and 
bound state nucleons, as discussed following Eq.\ \ref{waveA}.
%A more detailed description of phase space occupation
%by free and bound nucleons can be given within the cluster mean-field
%approximation, see Ref.\ \cite{Ropke:2008qk}.
It is clearly seen that the weakly bound deuteron dissolves in the medium at 
much lower densities than the more tightly bound $\alpha$-particle.

The quadratic form (\ref{eq:DBq}) of the binding energy shift 
$\Delta B_{i}$ predicts a transition of the cluster bound state
to the continuum at a transition density of
%\begin{equation}
$ \tilde{n}_{i}^{t}(T) = \left( \sqrt{3}-1\right) \tilde{n}_{i}^{0}(T) $
%\end{equation}
where the cluster binding energy $B_{i}$ % = B_{i}^{0}+\Delta B_{i}$ 
becomes zero, i.e.
there are only scattering correlations remaining and the energy 
corresponds to that of a resonance. In case of the
triton, helion and $\alpha$-particle, the resonance energy rapidly
moves to larger energies in the continuum with increasing
density leading to a strong suppression of the cluster fractions.
In contrast, the deuteron-like resonance stays closer to the
threshold causing a much weaker suppression of two-particle
correlations at high densities. 
The change of the binding energy from positive to
negative values allows to decribe
a continuous suppression of the cluster fraction 
with increasing density. A simple neglection of the cluster
contribution to the EoS as soon the energy crosses zero would
lead to an unphysical jump in the cluster density and the
thermodynamical properties.
In the QS model, the bound state contributions are 
rapidly cancelled by the continuum contributions, 
Eqs.\ (\ref{quasigas2_p},\ref{quasigas2_n}), leading to a
more rapid suppression 
also of the deuteron correlations. The differences in the continuum 
correlations will be seen to have large effects in the comparison 
of the results from the RMF and QS models.

\subsection{Model parameters}
\label{subsec:mp}

The generalized RMF model contains several parameters: the masses of
the particles, the couplings and binding energy shifts with their
specific functional dependence on densities and temperature.
In the present approach, experimental neutron and
proton masses $m_{n}$ and $m_{p}$ are used instead of an average
nucleon mass $m_{\rm nuc} = (m_{n}+m_{p})/2$. With the experimental
binding energies $B_{i}^{0}$ in the vacuum from \cite{Audi:2002rp}
the masses (\ref{eq:mass_cl}) 
of the clusters are also fixed. For the masses of the
$\omega$ and $\rho$ meson standard values of previous RMF models
are assumed. The mass of the $\sigma$ meson is determined from a
fit of the RMF parameters to properties of finite nuclei (see below).
The numerical values of the nucleon and meson masses are given
in table \ref{tab:02}.

\begin{table}[t]
\caption{\label{tab:02}%
Masses of the nucleons and mesons in the relativistic mean-field model.}
\begin{tabular}{cccccc}
\hline \hline
 particle $i$ & neutron & proton & 
 $\omega$ meson & $\sigma$ meson & $\rho$ meson \\
\hline
 $m_{i}$ [MeV] & 939.56536 & 938.27203 & 
 783 & 546.212459 & 763 \\
\hline \hline
\end{tabular}
\end{table}

The functional dependence of the couplings on the density is described by 
\begin{equation}
 \Gamma_{i}(n) = \Gamma_{i}(n_{\rm sat}) f_{i}(x)
\end{equation}
with $x=n/n_{\rm sat}$ where a rational function
\begin{equation}
 f_{i}(x) = a_{i} \frac{1+b_{i}(x+d_{i})^{2}}{1+c_{i}(x+d_{i})^{2}}
\end{equation}
is used
for the isoscalar mesons $i=\omega,\sigma$ and an exponential function
\begin{equation}
 f_{i}(x) = \exp[-a_{i}(x-1)]
\end{equation}
for the isovector meson $i=\rho$. In order to reduce the number of 
independent parameters the conditions $f_{i}(1) = 1$ 
and $f_{i}^{\prime\prime}(0)=0$ are imposed on the rational function.
These functions were introduced in \cite{Typel:2005ba} and are now widely used 
in RMF models with density-dependent couplings. 
The saturation density $n_{\rm sat}$, the mass of the $\sigma$ meson 
$m_{\sigma}$, 
the couplings $\Gamma_{i}(n_{\rm sat})$ and the
coefficients $a_{i}$, $b_{i}$, $c_{i}$ and $d_{i}$ are found by
fitting the properties of finite nuclei (binding energies, spin-orbit
splittings, charge and diffraction radii, surface thicknesses and
the neutron skin thickness of ${}^{208}$Pb) in the same way as for
the parametrization DD in Ref.\ \cite{Typel:2005ba}.  
In total, there are ten independent parameters in the fit. 
Numerical values of the coupling parameters can be found in table \ref{tab:03}.
This new parametrization is called DD2 since it is a modification of the set
DD where the only difference is the use of experimental nucleon masses.

\begin{table}[t]
\caption{\label{tab:03}%
Parameters of the couplings in the relativistic mean-field model.}
\begin{tabular}{cccccc}
\hline \hline
 meson $i$ & $\Gamma_{i}(n_{\rm sat})$ & 
 $a_{i}$ & $b_{i}$ & $c_{i}$ & $d_{i}$ \\
 \hline
 $\omega$  & 13.342362 & 1.369718 & 0.496475 & 0.817753 & 0.638452 \\
 $\sigma$  & 10.686681 & 1.357630 & 0.634442 & 1.005358 & 0.575810 \\
 $\rho$    &  3.626940 & 0.518903 &          &          & \\
\hline \hline
\end{tabular}
\end{table}

On the basis of this fit, the saturation density of symmetric nuclear
matter at zero temperature
is obtained as $n_{\rm sat} = 0.149065$~fm${}^{-3}$ with a binding energy per 
nucleon of $-16.02$~MeV. 
The incompressibility turns out to be $K_{\infty}=242.7$~MeV with a derivative 
$K^{\prime} = -529.8$~MeV.
See \cite{Typel:2005ba,Klahn:2006ir} for the definition of these quantities. 
These values are very reasonable and close to the results of other modern RMF
parametrizations. 
The large negative value of $K^{\prime}$ is a result of the fit to the surface 
properties of nuclei and leads a rather stiff EoS for symmetric 
nuclear matter at high densities. 
The small effective Dirac mass at saturation of $0.5625~m_{\rm nuc}$ is 
required in order to get a good description of the spin-orbit
splittings. This corresponds to an effective Landau mass of
$m^{\ast}= 0.6255~m_{\rm nuc}$.

The value and the density dependence of the symmetry energy near the nuclear
saturation density $n_{\rm sat}$ are usually characterized by the quantities 
$J = E_{\rm sym}(n_{\rm sat},0,0)$ and
the slope parameter
$L = 3 \left. dE_{\rm sym}/dn \right|_{n=n_{\rm sat},T=0}$.
%and the symmetry incompressibility
%$K_{\rm sym} = 18 \left. d^{2}E_{\rm sym}/dn^{2} \right|_{n=n_{\rm
%    sat}, T=0}$.
With the parametrization of the present RMF model, the values
$J=32.73$~MeV and
$L=57.94$~MeV %and $K_{\rm sym}=-93.49$~MeV 
are found. The obtained symmetry
energy at saturation $J$ is fully consistent with all modern RMF 
parametrizations and expectations. The rather small slope coefficient
$L$ is a consequence of fitting the neutron skin thickness of ${}^{208}$Pb
that is not known precisely so far. Similar low values for $L$ are found in
other contemporary 
RMF parametrizations with density dependent couplings or with
extended non-linear meson self-interactions. Older non-linear RMF models
were not able to give a reasonable value of the neutron skin thickness
with values for $L$ in excess of 100~MeV. They displayed a much
stiffer symmetry energy because of a restricted form
for the isospin dependence of the interaction.
The symmetry energy of nucler matter parameters at subsaturation densities 
is one of the central results of this work and will be discussed in section 
\ref{sec:SymE}.

\section{Properties of symmetric nuclear matter with light clusters}
\label{sec:SNM}

The appearance of light clusters in nuclear matter at densities below
saturation affects the composition and the thermodynamical 
properties
of the system. In this section we will compare the results of the
quantum statistical approach (QS) with the generalized relativistic 
mean-field model (RMF) 
in reference to the nuclear statistical equilibrium (NSE) model, 
which gives the correct
behavior in the limit of small densities. 
We will start to discuss the composition of the system, which shows most
directly the differences of the models.
In all figures in this
section we consider isothermes in symmetric nuclear matter
as a function of the total baryon
density for temperatures between 2~MeV and 20~MeV in steps of 2~MeV
keeping the same color code. This
representation immediately allows to study 
the systematic evolution of the various properties.

\subsection{Composition}
\label{subsec:comp}

\begin{figure}
\epsfig{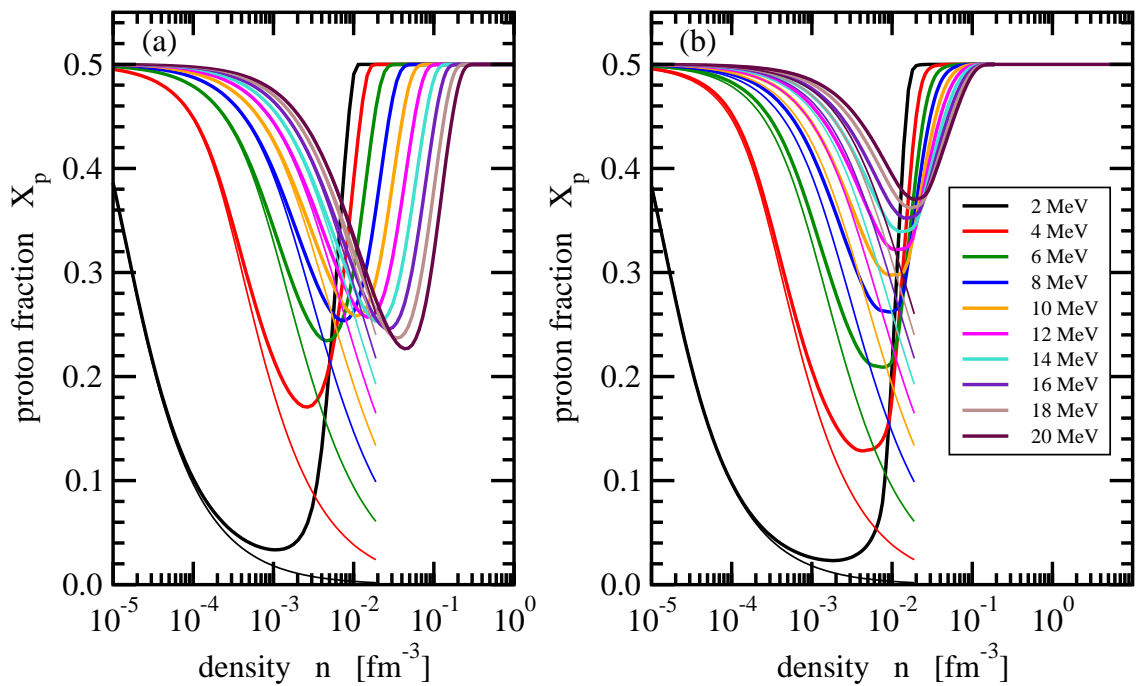}
\caption{\label{fig:07}%
(Color online)
Proton fraction $X_{p}$
in symmetric nuclear matter
as a function of the total density $n$ 
for various temperatures $T$
in the generalized RMF model (a) and
the QS approach (b) with light clusters (thick solid lines).
The result of the NSE calculation with light clusters is denoted
by thin solid lines for low densities. }
\end{figure}

We start to discuss the composition of the system, 
i.e.\ the 
fractions $X_{i} = A_{i} n_{i}/n$ of the various particle species $i$
as calculated in the RMF and the QS models.
We first describe the results and then attempt to give an explanation 
of the differences between the models at the end of the subsection.

In Fig.\ \ref{fig:07} the density dependence of the free proton fraction,
i.e. of protons not bound in a cluster,
is shown for different temperatures in symmetric nuclear matter
(which in this case is nearly identical to the neutron fraction).
The two approaches are compared to the NSE model
up to densities $n < 2 \cdot 10^{-2}$~fm${}^{-3}$. 
For finite temperatures $T$ the free proton fraction
in symmetric nuclear matter always approaches the value $0.5$  for
$n \to 0$~fm${}^{-3}$.  
It first decreases with increasing densities because of the
formation of clusters, but then increases again because the clusters 
dissolve at higher densities. Eventually the system becomes homogeneous 
again, and the nucleon fractions attain the value 0.5.
For densities below $n \sim 10^{-4}$~fm$^{-3}$ the
fraction of free protons in both models is very well described by the
NSE result since here mean-field effects and changes of the cluster
properties are practically negligible. With increasing density, the NSE
proton fraction approaches zero asymptotically irrespective of
the temperature, i.e.\ all
protons are predicted to be bound in clusters. This unphysical result
does not occur in both the RMF and QS approaches.
Instead, the clusters dissolve at high densities leading to free
protons and neutrons at high densities, i.e.\ the correct limit
is obtained.

The two approaches, RMF and QS, generally show a similar behavior.
However, in the transition region where the clusters dissolve,
there are significant differences, which are
more pronounced at high temperatures.
In the QS approach the minimum of the free proton fraction increases
with the temperature monotonously, and the
minimum position is slightly shifted to higher values in density
but stays close to $10^{-2}$~fm${}^{-3}$.
Essentially all protons are free at saturation density 
independent of temperature.
In the RMF model the behavior at temperatures
above $\approx 8$~MeV is different. The minimum of $X_{p}$
starts to decrease again and the minimum position
moves considerably to higher densities with increasing temperature, 
such that the model predicts that some protons are still bound
in clusters even at saturation density. 
%A closer inspection of the
%cluster fractions will give a hint for the origin of this behavior. 
%This will also explain the noticeable difference between the NSE results and
%the QS predictions for $X_{p}$ at temperatures larger than 2~MeV.

\begin{figure}
\epsfig{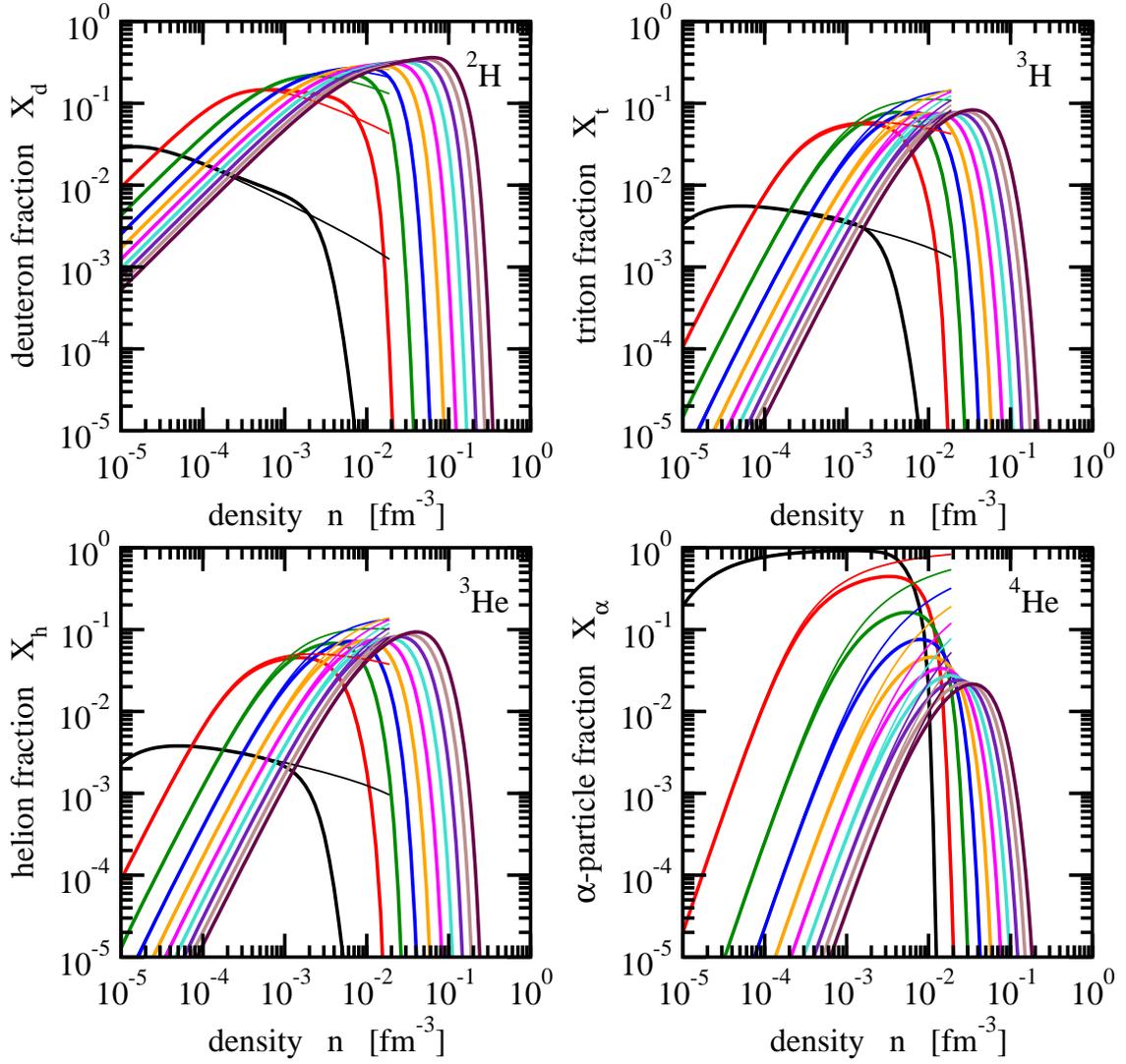}
\caption{\label{fig:08a}%
(Color online)
Cluster fractions $X_{i}$ in symmetric nuclear matter
as a function of the total density $n$ 
for various temperatures $T$
in the generalized RMF model (thick solid lines).
The result of the NSE calculation with light clusters is denoted
by thin solid lines for low densities. See Fig.~\ref{fig:07}
for the color code.}
\end{figure}

\begin{figure}
\epsfig{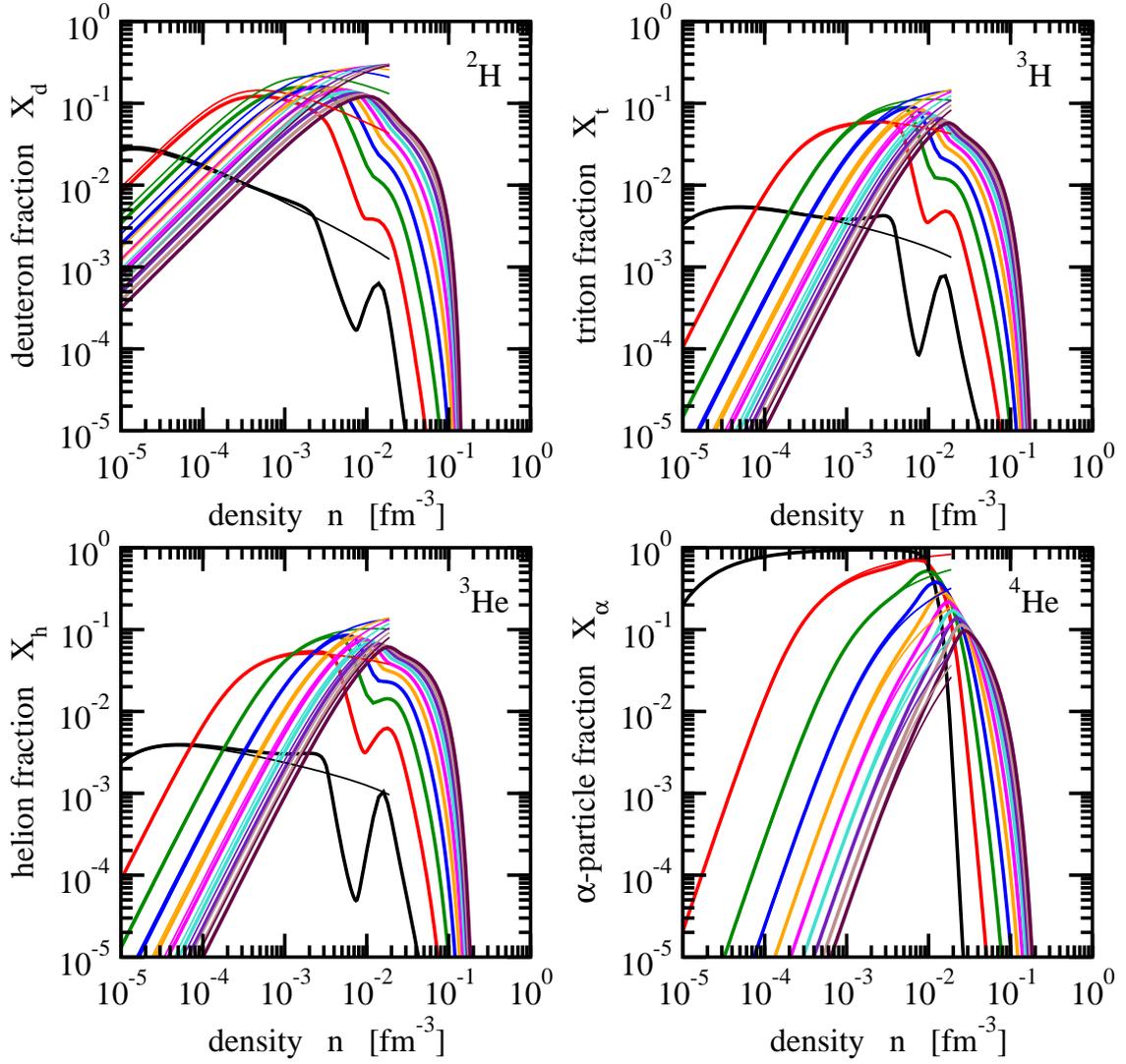}
\caption{\label{fig:08b}%
(Color online)
Cluster fractions $X_{i}$ in symmetric nuclear matter
as a function of the total density $n$ 
for various temperatures $T$
in the quantum statistical approach (thick solid lines).
The result of the NSE calculation with light clusters is denoted
by thin solid lines for low densities. See Fig.~\ref{fig:07}
for the color code.}
\end{figure}

The information on the fractions of the different light
clusters, deuteron to $^4$He, is shown
in Figs.\ \ref{fig:08a} for the
RMF model and in Fig.\ \ref{fig:08b} for the QS model, 
again in comparison with NSE.
At zero temperature and density the cluster fractions are simply 
determined by the binding energies. Thus in symmetric nuclear matter 
all nucleons are bound in alpha particles. 
With increasing temperature the cluster fractions
show the complementary behavior compared to the free nucleon fractions, 
discussed above. 
At low densities and higher temperatures all cluster fractions are small 
decreasing with temperature but increasing with
density. For a fixed temperature 
first the deuteron appears, then the three-body bound
states and finally the $\alpha$ particle. The distribution between
the clusters changes with increasing density of the medium
such that correlations between more and more nucleons become important.
As a consequence, the fractions of the lighter clusters 
decrease whereas the $\alpha$-particle fraction
still increases. 
In the QS model (but not in the RMF model) at higher
densities the more strongly bound
$\alpha$-particle are formed more frequently than deuterons.
The same trend is observed for the NSE calculation.
However, for densities beyond nuclear saturation, the NSE predicts that
all nucleons in symmetric
nuclear matter would be bound in clusters, which again is unphysical
since in this model the medium modification and the eventual 
disappearance of the clusters is not taken into account.

Again there are significant differences in the two approaches.
In the RMF model the cluster fractions along isothermes rise and fall
monotonously with a single maximum. 
In general they stay below the fractions predicted by
the NSE calculation, since the clusters are less bound 
inside the medium.
In the QS approach, on the other hand, at lower temperatures, 
the fractions of deuterons, tritons and 
helions exhibit a sudden drop around densities 
of $10^{-2}$~fm${}^{-3}$, which is accompanied by an increase in the
$\alpha$-particle fraction as compared to the NSE result. 
Generally, the range of densities where the clusters disappear is more confined
in the QS relative to the RMF model.
%Especially
%at high temperatures, the latter model predicts a much higher
%deuteron fraction in the transition region from clusterized matter
%to pure nucleon matter. 

Many of the differences in the behavior of the models can be 
traced back to the different treatment of the deuteron correlations. 
The QS approach takes the continuum contributions explicitly into
account, which effectively reduces the strength of the two-body correlations,
as is, e.g., seen in Eqs.\ (\ref{quasigas2_p}), (\ref{quasigas2_n}).
In RMF, on the other hand, the deuteron correlations are represented 
by a single state that slowly moves to higher
energies in the continuum, and thus two-body correlations are overestimated.
Also in NSE the deuteron correlations are overestimated, 
since there is no continuum contribution.
This has an strong effect on the deuteron fraction 
and - in competition - also on the fractions of nucleons and other clusters. 
We see that in the RMF model the deuteron fractions 
at higher densities are generally larger than
in the QS model, as seen by comparing the upper left panels of 
Figs.\ \ref{fig:08a} and \ref{fig:08b}. 
This, in turn, has a strong effect on the 
$\alpha$ fractions, which are much lower in the RMF model as seen in the 
lower right panels of these figures, but also
leads to the decrease of the free nucleon fraction at higher
densities, 
as seen in Fig.\ \ref{fig:07}. 
These effects can also be seen in the comparison with the NSE limit, 
where they lead to larger deuteron fractions relative to the QS model. 
Only at temperatures smaller than the deuteron binding energy, 
the bound state is the dominating two-body correlation.
A similar effect occurs in the virial
description of matter at low densities that is encoded in the
temperature denpendence of the second virial coefficient.
For heavier clusters the influence of the continuum on the fractions
is much less pronounced due to their larger binding energies.

As we remarked above, the QS model shows a particularly enhanced alpha particle
fraction at the higher densities, as seen in the increase above the NSE limit 
at densities around $n \sim 10^{-2}$~fm$^{-3}$ in Fig.\ \ref{fig:08b}. 
This has much of the appearance of an onset of an alpha particle condensation. 
As a consequence, the fractions of the other clusters show a dip around this 
density, and there are also consequences in the thermodynamical quantities, as 
seen below. 
All these effects are not present in the RMF model. 
However, mean field contributions from the rather substantial cluster fractions
in this density range are not taken into account in the QS model, the effect of
which needs to be further investigated. 
It is important to note that the $\alpha$ particle is usually not the
most frequent cluster and that there are substantial contributions
from the deuteron, the triton and the helion at intermediate densities
and temperatures that are not considered in the EoS 
of Lattimer and Swesty \cite{Lattimer:1991nc} or
Shen, Toki et al.\ \cite{Shen:1998gq}. Furthermore,  
the excluded-volume mechanism to suppress the formation of clusters
at high densities does not take into account any temperature
dependence in this process which is clearly present in our more
microscopic models.

\subsection{Thermodynamical quantities}
\label{subsec:tq}

\begin{figure}
\epsfig{file=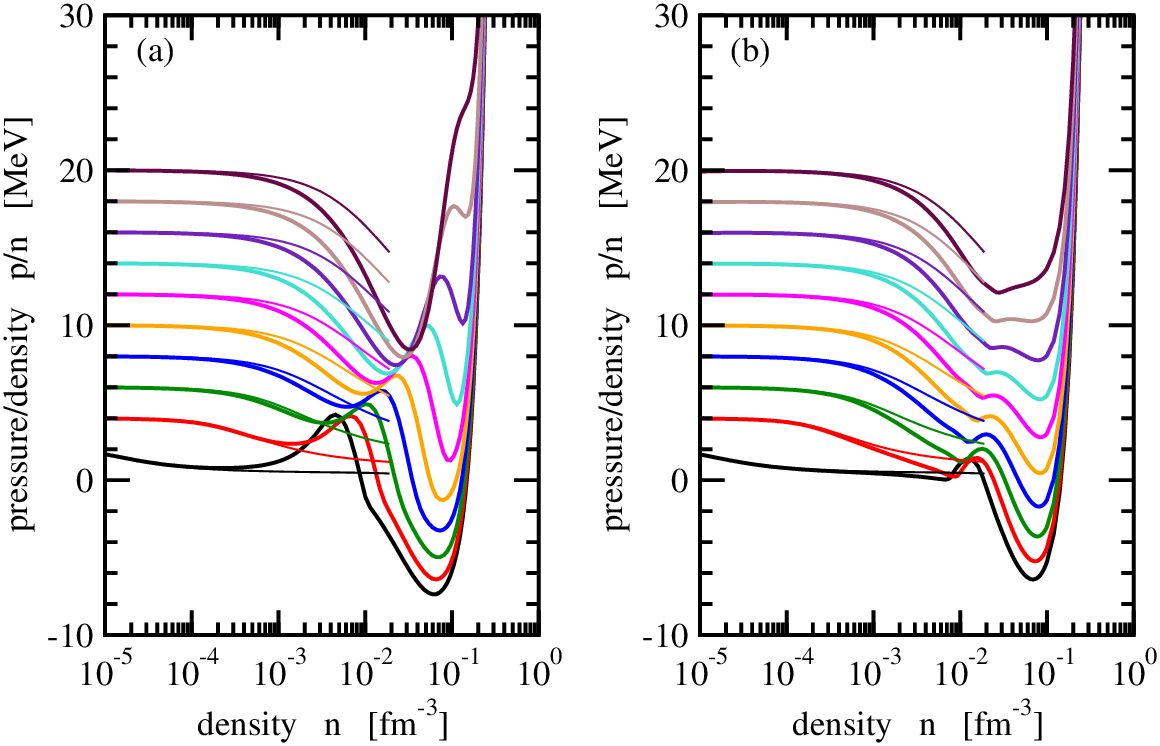,width=15cm}
\caption{\label{fig:01}%
(Color online)
Ratio pressure over density $p/n$ of symmetric nuclear matter
as a function of the total density $n$ 
for various temperatures $T$ in the generalized RMF model (a) and
the QS approach (b) with light clusters (thick solid lines).
The result of the NSE calculation with light clusters is denoted
by thin solid lines for low densities.  See Fig.~\ref{fig:07} for the 
color code.}
\end{figure}

In this subsection we discuss the thermodynamical quantities 
for symmetric nuclear matter.
The essential effects of the formation of clusters in the various
models are given by the 
pressure $p$ as a function of the
total baryon number density density $n$.
For a better representation, we depict in Fig.\
\ref{fig:01} the ratio $p/n$ as a function of $n$. 
The left  and right panels of the figure show the results of the
RMF and the QS models, respectively, with thick lines. The thin lines in both
panels represent a NSE calculation
with neutrons, protons and the light clusters $d$, $t$, $h$ and
$\alpha$. They are shown only for densities
below $2 \cdot 10^{-2}$~fm${}^{-3}$ because
the contribution of heavier clusters can be substantial at higher
densities, at least for low temperatures, and the NSE becomes
unrealistic.

At very low densities, all models approach the ideal gas limit
with $p/n=T$ since the cluster fraction is very small and nuclear
matter is composed primarily of neutrons and protons (see Subsect.\
\ref{subsec:comp}). 
Relativistic and
mean field effects are not important here. With increasing density the NSE
calculation exhibits a reduction of the pressure relative to the
ideal gas that is caused by the formation of light clusters. Both
the RMF and the QS models follow this trend but there are
considerable differences.
The QS results stays closer to the NSE calculation whereas 
RMF shows a deviation already at lower densities
around $10^{-3}$~fm. However, both models predict a stronger
decrease of the pressure than the NSE, because at
these densities the changes of the cluster properties and the
mean-field effects are already effective.
An exception is seen for
the RMF approach at temperatures below approx.\ 6~MeV
where it shows an increase relative to the NSE.

In the density range from $10^{-2}$ to 
$2 \cdot 10^{-1}$~fm${}^{-3}$ the ratio $p/n$ 
displays considerable structure and passes through
at least one minimum, eventually with negative pressure, 
before it rises sharply for densities above nuclear
saturation ($p/n=0$) and the matter becomes very 
incompressible. In this region, we observed the most 
pronounced differences in the  behavior of the
RMF and QS models with respect to the composition in Subsect.\ 
\ref{subsec:comp}. 
They are generated partly by differences in the description of the cluster 
correlations
and partly by a different treatment of the mean-field effects.
In the QS approach, the strengths of the mean fields and thus
the quasi-particle energies and mass shifts are independent
of the composition of the system since they are taken 
in parametrized form from
the RMF calculation without clusters.
In contrast, in the RMF approach 
the additional contributions in the source terms
of the meson fields, see Eqs.\ (\ref{eq:omega}) and (\ref{eq:rho}),
depend on the cluster densities. They describe the back reaction
of the cluster formation on the medium (apart from the
contribution of the bound nucleons) and are inevitable for the
thermodynamical consistency of this model. In the density region where
the cluster fraction is substantial, the additional terms lead to
a sizable modification of the vector meson fields.
In the density range around $n \sim 10^{-2}$~fm${}^{-3}$ the pressure is 
lower in the QS relative to the RMF model. 
This correlates with increased alpha 
formation ($\alpha$ ``condensation'') in this model relative to RMF, 
as was seen in Fig.\ \ref{fig:08b}.

\begin{figure}
\epsfig{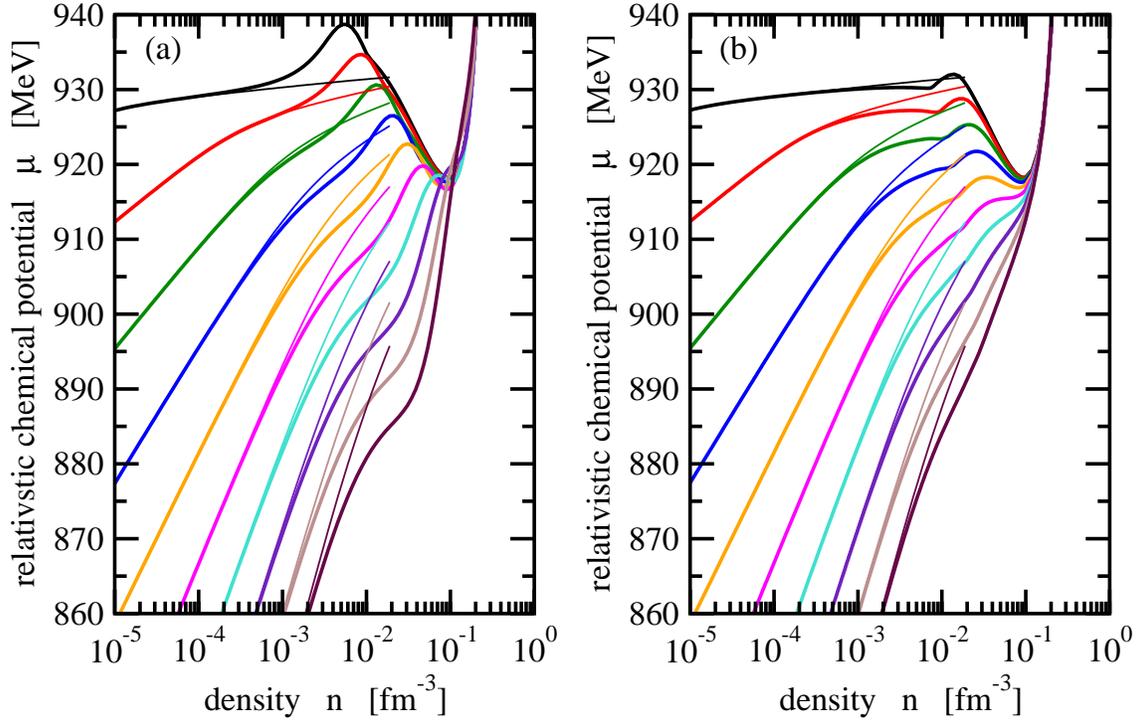}
\caption{\label{fig:02}%
(Color online)
Relativistic baryon chemical potential $\mu=(\mu_{p}+\mu_{n})/2$ 
of symmetric nuclear matter
as a function of the total density $n$ 
for various temperatures $T$
in the generalized RMF model (a) and
the QS approach (b) with light clusters (thick solid lines).
The result of the NSE calculation with light clusters is denoted
by thin solid lines for low densities. See Fig.~\ref{fig:07}
for the color code.}
\end{figure}

The relativistic baryon chemical potentials
$\mu = (\mu_{p}+\mu_{n})/2$ 
of the RMF and the QS approaches are displayed
in Fig.\ \ref{fig:02} in comparision with the NSE calculation.
They reflect the behavior observed for the pressure of the various 
models.
At very low densities, the three calculations agree with each other
and the ideal gas dependence of the chemical potential
$\mu = m + T \ln \left( n\lambda^{3}/4\right)$ on temperature and density.
With increasing density, the chemical potential of both the
RMF and QS calculations are below the NSE results,
except for the RMF model at low temperatures 
as already seen for the pressure.
Here, the chemical
potential rises above the NSE result with increasing
density and a pronounced maximum occurs. In the QS model, the
chemical potential generally stays below the NSE result.
Deviations from the NSE predictions appear already at lower densities
in the RMF relative to the QS model.
In general, however, the differences between two approaches
are less obvious in the chemical potential than in the
pressure. In both models the low-density and the high-density
limit are correctly described.

\begin{figure}
\epsfig{file=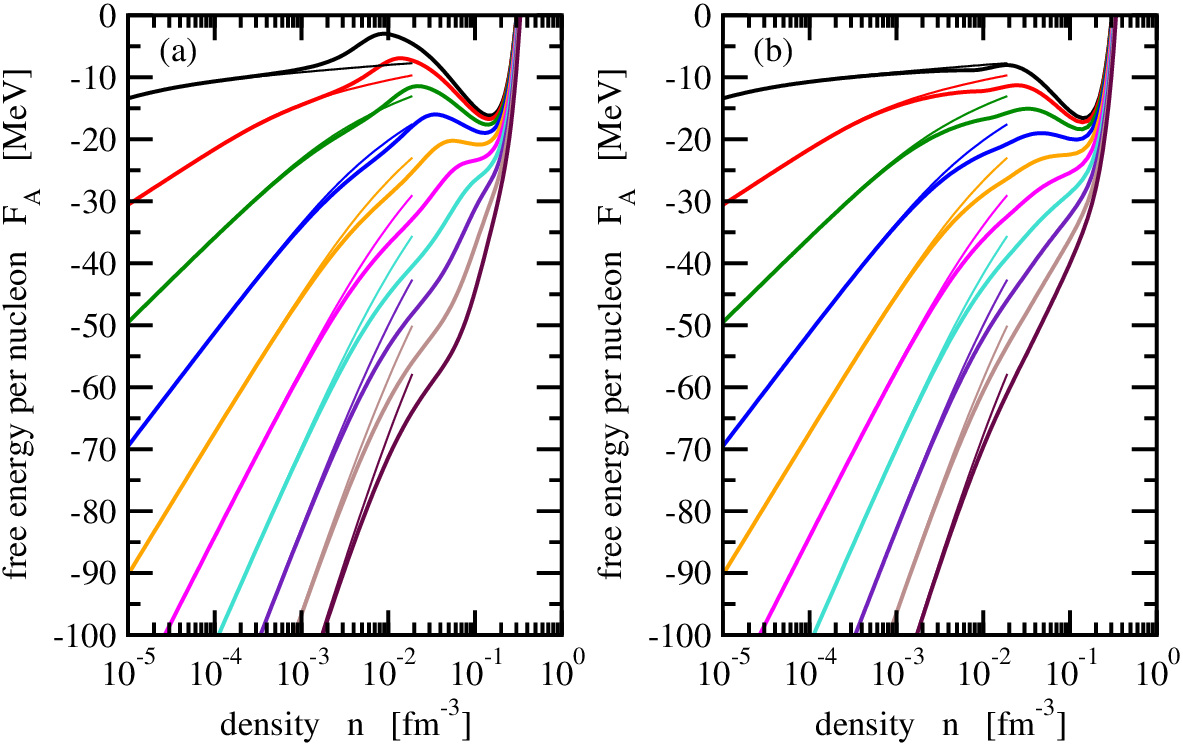,width=15cm}
\caption{\label{fig:04}%
(Color online)
Free energy per nucleon $F_{A}$ 
of symmetric nuclear matter
as a function of the total density $n$ 
for various temperatures $T$
in the generalized RMF model (a) and
the QS approach (b) with light clusters (thick solid lines).
The result of the NSE calculation with light clusters is denoted
by thin solid lines for low densities. See Fig.~\ref{fig:07}
for the color code.}
\end{figure}

The density dependence of the free energy per nucleon $F_{A}$,
Eq.\ (\ref{eq:FA}), is shown in Fig.\ \ref{fig:04} and closely
follows the density dependence of the chemical potential. 
In the QS approach, $F_{A}$ is obtained by a simple direct integration
of the chemical potential, cf.\ Eq.\ (\ref{eq:freV}). 
Both models agree perfectly
with the NSE result for densities below $10^{-4}$~fm${}^{-3}$.
At the nuclear saturation density, the free binding energy per nucleon
approaches the local minumum at $-16$~MeV 
in the limit $T \to 0$~MeV as expected.

\begin{figure}
\epsfig{file=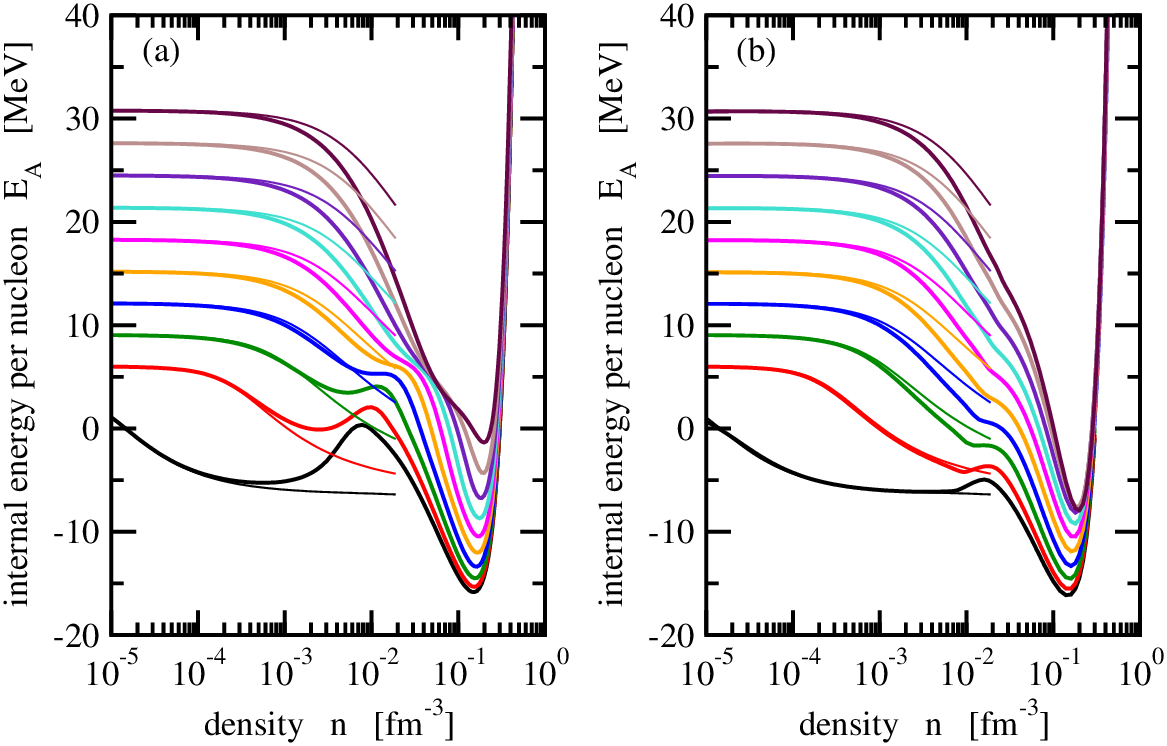,width=15cm}
\caption{\label{fig:05}%
(Color online)
Internal energy per nucleon $E_{A}$ 
of symmetric nuclear matter
as a function of the total density $n$ 
for various temperatures $T$
in the generalized RMF model (a) and
the QS approach (b) with light clusters (thick solid lines).
The result of the NSE calculation with light clusters is denoted
by thin solid lines for low densities. See Fig.~\ref{fig:07}
for the color code.}
\end{figure}

The results for the internal energy per nucleon
$E_{A}$ are depicted in Fig.\ \ref{fig:05}.  
Here the differences between
the RMF and QS results are less pronounced than
in the chemical potential $\mu$ or the free binding energy per nucleon 
$F_{A}$. In contrast to the latter quantities, $E_{A}$
increases with temperature at a given density.
For $n \to 0$ the classical limit
$E_{A} \to 3T/2$ of an ideal gas is approached.
Slight deviations stem from the use of relativistic dispersion relations.

\begin{figure}
\epsfig{file=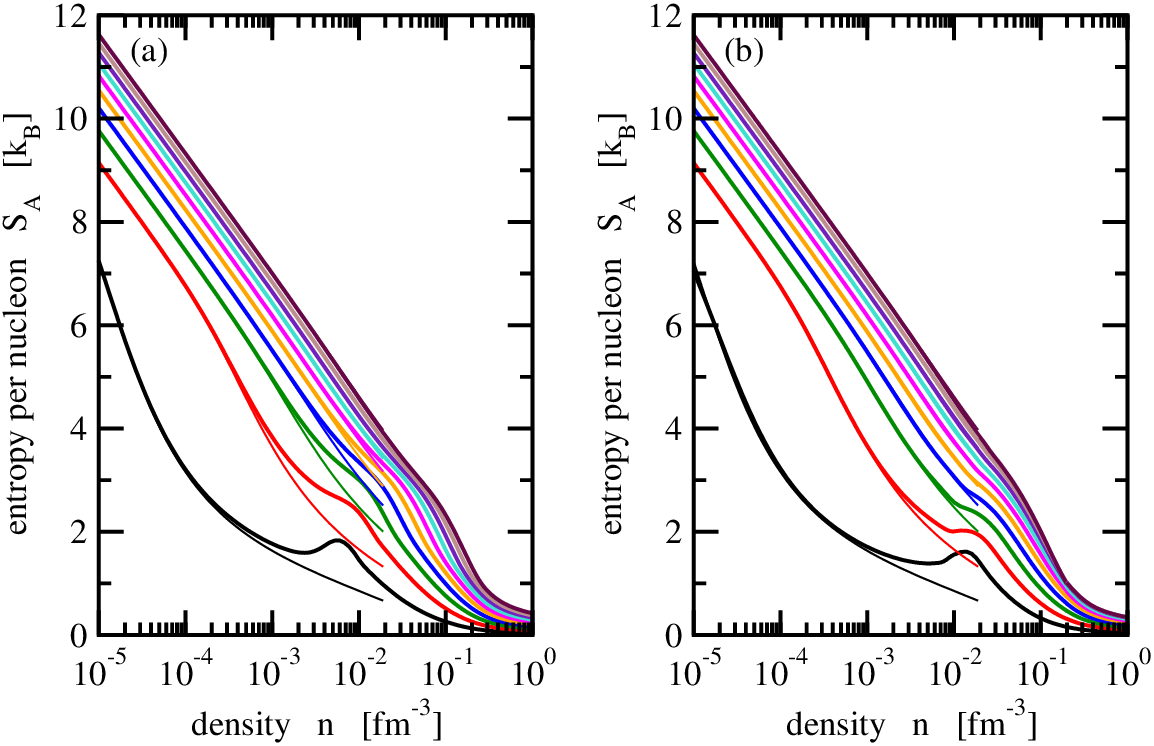,width=15cm}
\caption{\label{fig:06}%
(Color online)
Entropy per nucleon $S_{A}$ 
of symmetric nuclear matter
as a function of the total density $n$ 
for various temperatures $T$
in the generalized RMF model (a) and
the QS approach (b) with light clusters (thick solid lines).
The result of the NSE calculation with light clusters is denoted
by thin solid lines for low densities. See Fig.~\ref{fig:07}
for the color code.}
\end{figure}

The entropy per nucleon $S_{A}=s/n$, shown in 
Fig.\ \ref{fig:06}, 
generally decreases monotonously
with increasing density $n$ except for a small
range in density at low temperatures when the light clusters
suddenly dissolve in the medium. At low densities the entropy per 
nucleon in the RMFand QS 
models approaches the NSE result and
at very low densities, a dependence $S_{A} \propto -\ln(n) + const.$
is found consistent with the behavior of an ideal gas.
The occurence of clusters is found to lead to a reduction of the
entropy per nucleon as compared to pure neutron-proton matter.

\section{Liquid-Gas Phase transition}
\label{sec:pt}

In a system of given total neutron and protons number densities,
$n_{n}^{\rm  tot}$ and $n_{p}^{\rm tot}$, and 
temperature $T$, the corresponding thermodynamical potential,
i.e.\ the free energy density $f(n_{n}^{\rm tot},n_{p}^{\rm tot},T)$, 
should be minimized.
However, this procedure does not necessarily give the correct
equilibrium state. 
Thermodynamic laws require that the free energy density is a convex function 
in the variables $n_{n}^{\rm tot}$, $n_{p}^{\rm tot}$ and $T$ to assure the 
stability of the system. 
This condition leads to the occurence of phase transitions. 
The coexistence region of different phases in thermodynamical equilibrium
is separated from the region of a single phase by the so-called
binodal surface. It can be obtained for a given temperature
by a general Gibbs construction
where the values of the intensive variables pressure $p$ and
the chemical potentials of protons $\mu_{p}$ and neutrons $\mu_{n}$
in the two phases have to be identical, see, e.g.\ \cite{Bar80,Mul95}.
Thus, the binodal surface is found from a global criterion
in contrast to the spinodal surface that defines the boundary
of local instability of the system, which occurs, e.g., when the 
compressibilities of the system become negative. 
In general, the spinodal is enclosed by the binodal and both become identical 
along critical lines. 
In the case of symmetric nuclear matter, the problem
becomes one-dimensional
and a usual Maxwell construction for the phase transition is
sufficient with constant pressure $p$ and baryon chemical potential $\mu$
for densities inside the coexistence region. 
In the general case of asymmetric matter, the pressure and chemical potentials 
do not stay constant. 
In addition, the low-density phase has a larger isospin asymmetry than the 
coexisting high-density phase.

Nuclear matter as considered here is an idealized physical system where the 
Coulomb interaction is neglected and charge neutrality is not demanded. 
The phase transition boundary as constructed by the
above standard procedure only gives a first indication of
where the formation of inhomogeneities and heavy nuclei occurs. 
In more realistic calculations with Coulomb interaction
the system has to be globally charge neutral and at least the contribution 
of electrons has to be considered, e.g.\ in applications of the EoS to 
astrophysics. 
We will not follow up on this issue in the present work.
For references to work in this direction, see our discussion of pasta phases
in the Introduction.

\begin{figure}
\epsfig{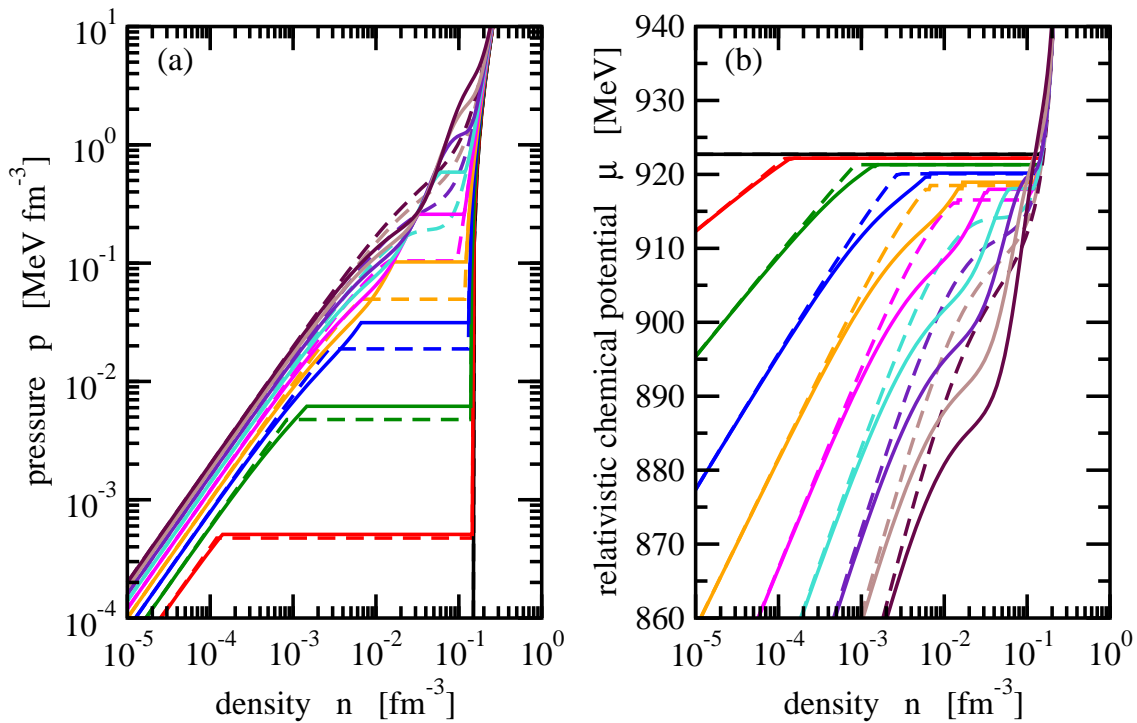}
\caption{\label{fig:03a}%
(Color online)
Pressure (a) and relativistic 
baryon chemical potential (b) of symmetric nuclear
matter as a function of the baryon number density in the
generalized RMF model with clusters (solid lines) and
without clusters (dashed line) taking the phase transition into account.
See Fig.~\ref{fig:07} for the color code.}
\end{figure}

The results for the pressure $p$, Fig.\ \ref{fig:01}, and the baryon chemical 
potential $\mu$, Fig.\ \ref{fig:02}, allow to construct the phase transition 
in symmetric nuclear matter from the gas phase with nucleons and light 
clusters at small densities to the liquid phase of pure nucleon matter at high 
densities.
In the following we consider the example of the RMF model.
In Fig.~\ref{fig:03a} the pressure $p$ and the chemical potential $\mu$ with 
phase transition are shown for the cases without and with light 
clusters. 
Inside the coexistence region of the two phases pressure and chemical potential
stay constant as characteristic for a Maxwell construction. 
As the temperature approaches zero, the density range of this region becomes 
larger with the limits $p \to 0$~MeV~fm${}^{-3}$ and $\mu \to m-16$~MeV.
The differences between the calculations without and with clusters
increase at higher temperatures with a larger coexistence pressure
in the latter case. Finally, the critical temperature is reached
beyond which no phase transition occurs any more.

At low temperatures the binodal line at the lower boundary 
of the coexistence region is reached already at very small densities.
At this point the cluster fraction is still
rather small, and, consequently, the appearance of clusters has little
effect on the determination of the phase boundary for low
temperatures. Larger effects of the clusters on the phase transition
are observed only at higher temperatures
where the cluster fraction at the 
lower density boundary reaches larger values. 
The binodal on the high density side of the phase coexistence
region is hardly affected when clusters are considered in the
calculation because here the matter is essentially composed of free
nucleons.

\begin{figure}
\epsfig{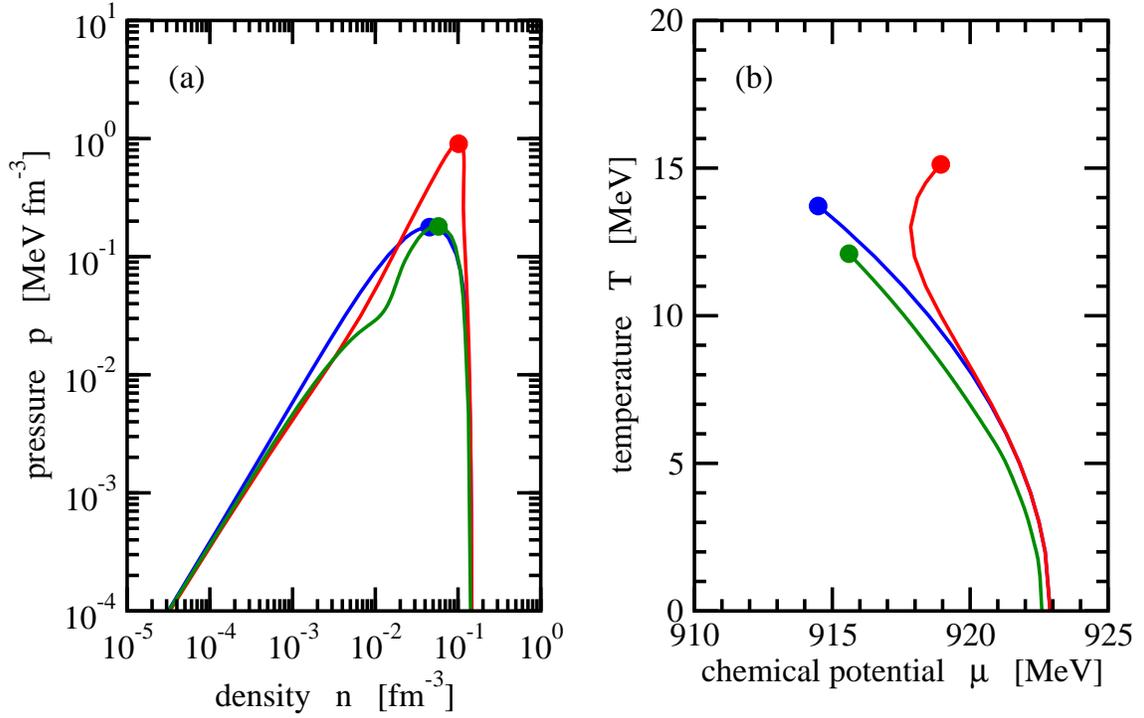}
\caption{\label{fig:03b}%
(Color online)
Binodal line enclosing the liquid-gas coexistence region 
in the pressure-density
diagram (a) and phase transition line in the temperature-chemical
potential diagram (b) for symmetric nuclear matter
in the RMF model without light clusters (blue lines),
with light clusters (red lines) and in the QS approach
(green lines).
The corresponding critical points are denoted by circles.}
\end{figure}

Results as in Fig.\ \ref{fig:03a}
can be used to extract the binodal line enclosing the
phase coexistence region in the $p-n$ diagram
and the phase transition line in the $T-\mu$ diagram. 
These are shown in the left and right parts of Fig.\ \ref{fig:03b},
respectively,
for the RMF model without and with clusters and the QS approach.
The occurence of light clusters in the system narrows the width of
the coexistence region and shifts the maximum to a higher density and
larger pressure in the generalized RMF model
as compared to the calculation without clusters.
In the RMF model the critical temperature $T_{c}$ 
increases from $13.72$~MeV without clusters to
$15.12$~MeV with clusters. The corresponding values of the
critical density $n_{c}$ are $0.0452$~fm$^{-3}$ and
$0.1018$~fm$^{-3}$, and of the critical pressure $p_{c}$
$0.1781$~MeV~fm$^{-3}$ and $0.9029$~MeV~fm$^{-3}$, respectively.
Thus, in the RMF model 
the position of the critical point shifts rather drastically
when light clusters are considered.
There is also a marked effect on the phase
transition line in the $T-\mu$ diagram.
In RMF without clusters, the chemical potential decreases monotonously 
with increasing temperature until it reaches the critial point
at a critical chemical potential of $\mu_{c} = 914.48$~MeV.
At low temperatures the phase transition line
with clusters follows closely the line without clusters. 
However, towards the critical point the transition line with light clusters 
has as $S$ shape.
At temperatures $T$ above $\approx 8$~MeV the curve finally bends 
to higher chemical potentials ending
at the critical point with $\mu_{c} = 918.9$~MeV.

In the QS approach a very different trend is found. 
The binodal line exhibits a characteristic dip on the
low-density side that is related to the enhancement of the
$\alpha$-particle fraction, cf.\ Fig.\ \ref{fig:08b}. The critical point
moves to a smaller temperature of $T_{c} = 12.1$~MeV and a slightly
higher chemical potential of $\mu_{c} = 915.61$~MeV as compared to
the RMF calculation without clusters. The critical pressure hardly
changes. The observed differences between the RMF
and the QS approaches are again related to the fact that many-body
correlations at high temperatures, especially of the deuteron,
are overestimated in the former. 
The behavior of the RMF approach must be considered unphysical 
in this respect, because one would expect, that correlations decrease 
the chemical potential and the pressure of the phase transition.
In an improved version of the generalized RMF approach
which takes the continuum contributions explicitly into account, 
it is expected that the critical point moves to lower temperatures.

\section{Symmetry energy}
\label{sec:SymE}

\begin{figure}
\epsfig{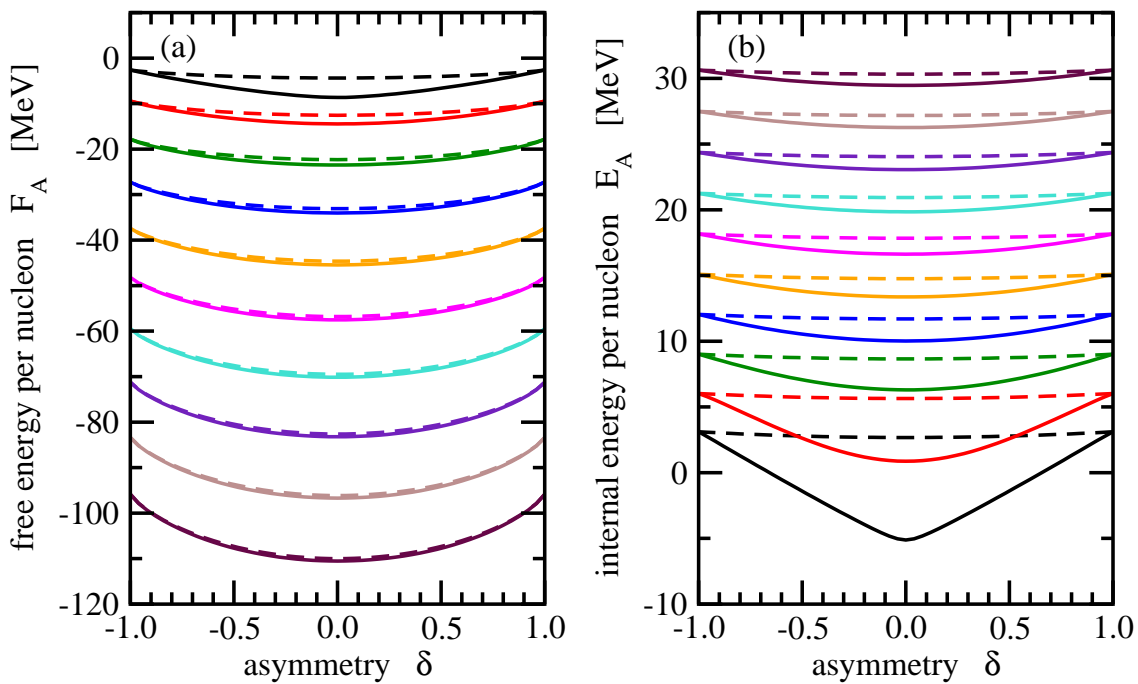}
\caption{\label{fig:09}%
(Color online)
Free energy per nucleon $F_{A}$ (left) and
internal energy per nucleon $E_{A}$ (right)
as a function of the asymmetry $\delta$ for constant
total density $n=0.001$~fm$^{-3}$
for various temperatures $T$ without light clusters (dashed lines) 
and with light clusters (solid lines).
See Fig~\ref{fig:07} for the color code.}
\end{figure}

The interanl energy per nucleon $E_{A}$ of asymmetric nuclear matter,
as defined in Eq. (\ref{eq:EA}), can be
expanded for given density $n$ and temperature $T$ 
in powers of the asymmetry $\delta$
\begin{equation}
 E_{A}(n,\delta,T) = E_{A}(n,0,T) + E_{\rm sym}(n,T) \delta^{2}
 + \dots \: .
\end{equation}
The symmetry energy is the coefficient of the first
term in the expansion that depends on the asymmetry $\delta$ quadratically.
Thus, it is defined as the second derivative
\begin{equation}
 \label{eq:esym}
 E_{\rm sym}(n,T) = \frac{1}{2} 
 \left. \frac{\partial^{2} E_{A}}{\partial \delta^{2}} 
 \right|_{\delta = 0} \: .
\end{equation}
Usually, the dependence of $E_{A}(n,\delta,T)$ 
for the complete range of asymmetries is
quite well approximated by a quadratic function. In this case
the symmetry energy also represents the difference between
the binding energy per nucleon of neutron matter and of symmetric nuclear
matter. At finite temperatures one has
to distinguish between the internal symmetry energy $E_{\rm sym}(n,T)$
and the free symmetry energy $F_{\rm sym}(n,T)$ that is similarly
defined.

However, the quadratic approximation
is not valid in general,
especially for a system with cluster correlations
at densities $n$ below the nuclear saturation density. 
This is clearly seen for our models in the dependence of $E_{A}$ and $F_{A}$
on $\delta$ for given $n$ and $T$.
In Fig.~\ref{fig:09} the free binding energy per nucleon (left)
and the internal binding energy per nucleon (right) are
shown as a function of the asymmetry for 
a fixed density $n=0.001$~fm$^{-3}$ and different temperatures. 
Without clusters, the dependence on $\delta$ 
is rather weak and can be well decribed by a parabolic function
irrespective of the temperature.
With clusters, the binding energies per nucleon are substantially
lowered around $\delta = 0$, i.e.\
symmetric nuclear matter, particularly for low temperatures. 
The system gains additional binding energy
by forming clusters.
Large deviations from a global quadratic dependence
develop and especially at low temperatures the parabola changes to
a triangular shape, most obviously seen for the internal energy
per nucleon. This behavior leads to a very large and thus not 
very meaningful symmetry 
energy in the limit $T \to 0$~MeV when the conventional definition 
(\ref{eq:esym}) is used. Hence, a more appropriate characterization
for the symmetry energy is required. A reasonable choice
is given by the finite difference formula
\begin{equation}
 E_{\rm sym}(n,T) = \frac{1}{2} \left[ E_{A}(n,1,T)
 - 2 E_{A}(n,0,T) + E_{A}(n,-1,T) \right]
\end{equation}
that is identical to (\ref{eq:esym}) for an exact quadratic dependence
of the internal binding energy per nucleon on $\delta$. 
This modified definition gives 
a good measure of the binding energy differences 
between neutron, proton, and symmetric nuclear matter. 
A corresponding equation defines the free symmetry energy $F_{\rm sym}(n,T)$.

\begin{figure}
\epsfig{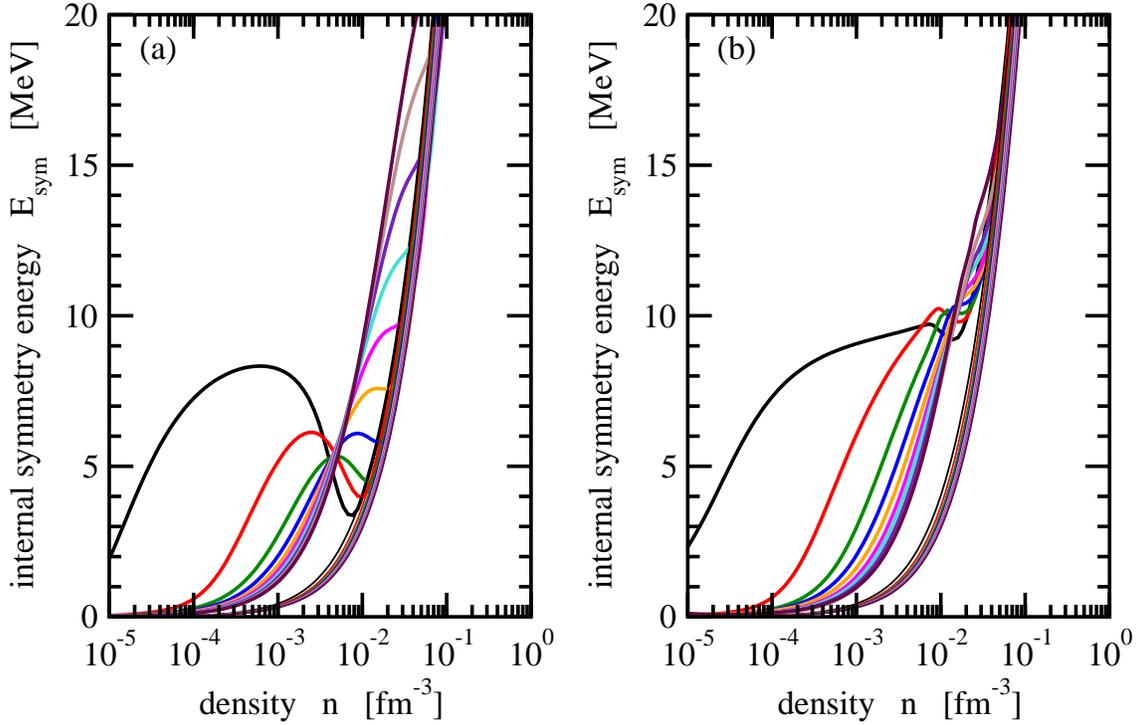}
\caption{\label{fig:10}%
(Color online)
Internal symmetry energy $E_{\rm sym}$ 
as a function of the total density $n$ 
for various temperatures $T$
in the generalized RMF model (a) and
the QS approach (b) with light clusters (thick solid lines).
The result of the RMF calculation without clusters is denoted
by thin solid lines for low densities. See Fig~\ref{fig:07}
for the color code.}
\end{figure}

\begin{figure}
\epsfig{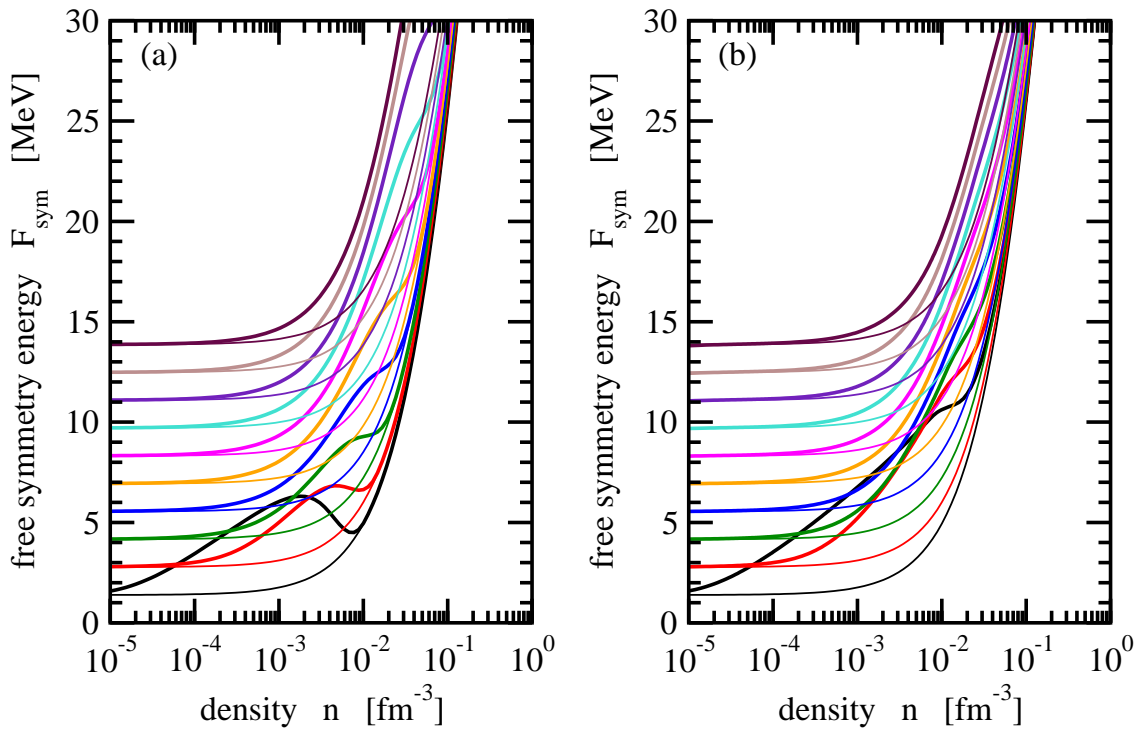}
\caption{\label{fig:11}%
(Color online)
Free symmetry energy $F_{\rm sym}$ 
as a function of the total density $n$ 
for various temperatures $T$
in the generalized RMF model (a) and
the QS approach (b) with light clusters (thick solid lines).
The result of the RMF calculation without clusters is denoted
by thin solid lines for low densities. See Fig.~\ref{fig:07}
for the color code.}
\end{figure}

The density dependence of the internal symmetry energy $E_{\rm sym}$
and of the free symmetry energy $F_{\rm sym}$ as defined above
is presented in Figs.\ \ref{fig:10} and \ref{fig:11}, respectively,
as a function of density for various temperatures.
In these figures, the results of the RMF 
and QS approaches with clusters are compared to
the RMF result without clusters. 
Just as for the 
free energy, Fig.\ \ref{fig:04}, and the internal energy, 
Fig.\ \ref{fig:05}, of symmetric nuclear matter,  the free and 
internal symmetry energies exhibit a different behavior for $n \to 0$.
In all models, $E_{\rm sym}$ approaches zero in this limit, but
$F_{\rm sym}$ converges to $T \ln 2$. 

In the RMF model without clusters both internal and free symmetry energies
rise continuously with increasing density, 
and $E_{\rm sym}$ is almost independent of $T$. 
When the formation of clusters is taken into account, the 
internal symmetry energy increases substantially at low 
densities, see Fig.\ \ref{fig:10}. This behavior is caused 
by the additional binding
of symmetric nuclear matter, that was already seen in  Fig.\ \ref{fig:09}, 
which is particularly pronounced at low
temperatures with a large cluster fraction. The density dependence
of the internal symmetry energy is rather different for the
for the RMF and QS approaches in a region near
$10^{-2}$~fm${}^{-3}$. As was discussed in Subsect.\ \ref{subsec:comp}
in the  RMF calculation the fraction
of deuterons is enhanced in the transition region whereas
three- and four-body correlations are suppressed as compared
to the QS approach. Correspondingly, symmetric nuclear matter is less
bound here and the symmetry energy is reduced relative to the QS approach.
Since two-body
correlations survive in the RMF model to too high
densities for larger temperatures, the internal symmetry energy
is also lowered considerably when the total baryon number density 
approaches the saturation density. In the QS approach, $E_{A}$
is much closer to the RMF result without clusters with
increasing density.
The free symmetry energy, see Fig.\ \ref{fig:11}, shows essentially
the same features as the internal symmetry. However, they are less
pronounced because of the different low-temperature limit.

\section{Comparison with other approaches}
\label{sec:alpha}

\begin{figure}
\begin{tabular}{cc}
\epsfig{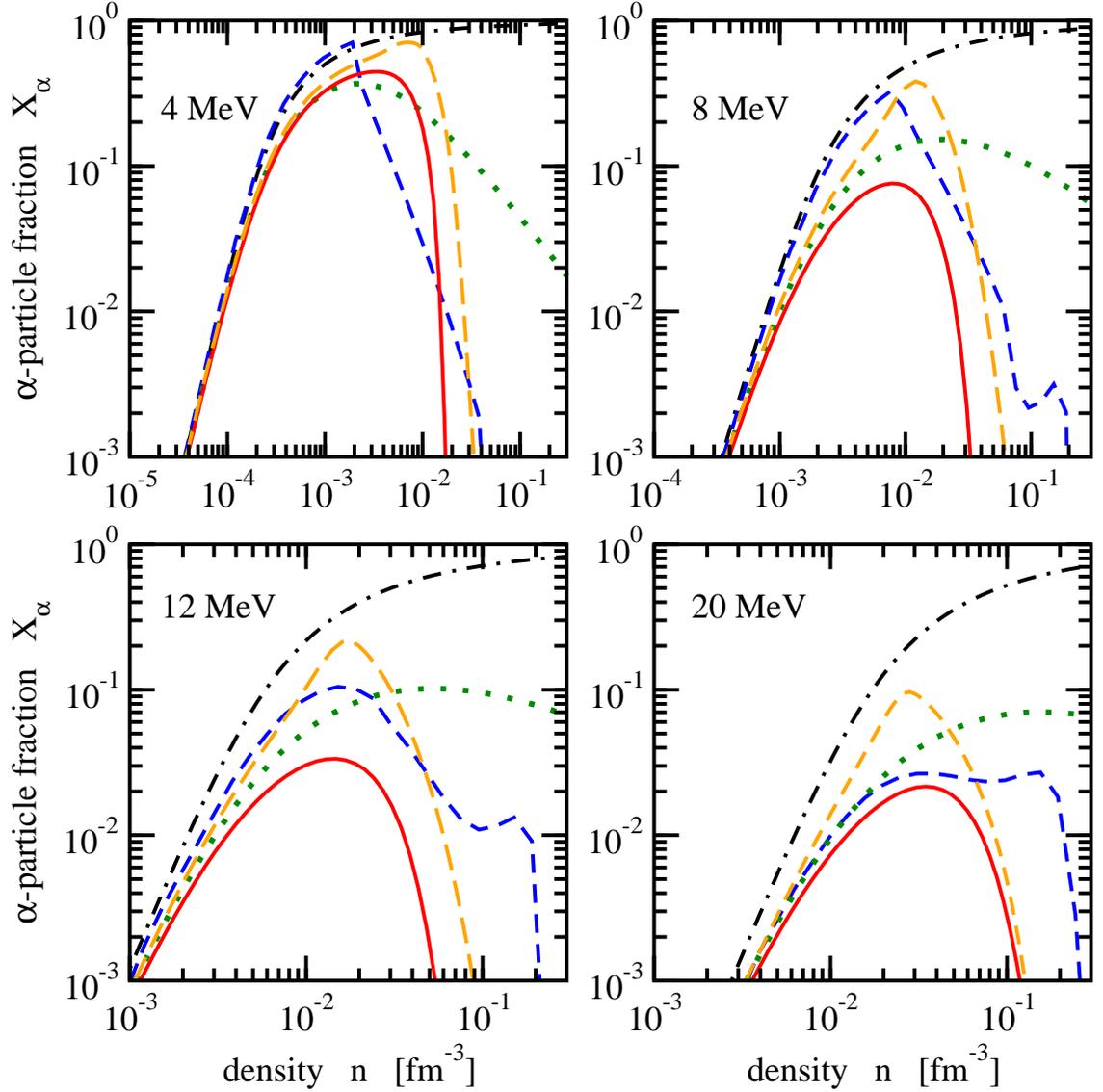}&
\end{tabular}
\caption{\label{fig:13}%
(Color online) 
Comparison of $\alpha$-particle fractions in symmetric nuclear matter 
as a function of the density at four temperatures for 
the virial expansion (black dashed-dotted lines), 
NSE (green dotted lines), the EoS of Shen et al. \cite{Shen:1998gq}
(blue dashed lines), the generalized RMF model (red solid lines) 
and the QS approach (orange dashed lines). Note the different scales
on the $x$-axes.}
\end{figure}

We discuss light cluster abundances obtained from our approaches 
in comparison with results from other models,
which appear in the literature and which have been discussed in this paper.
These are: the EoS of Shen et al.\ \cite{Shen:1998gq},
an EoS based on a virial expansion as presented in
\cite{Horowitz:2005nd},
and a nuclear statistical equilibrium calculation that takes into
account the ground states of all nuclei in the atomic mass evaluation
(AME 2003) \cite{Audi:2002rp}.
Fig.\ \ref{fig:13} shows the $\alpha$-particle fraction 
in symmetric nuclear matter  at four different temperatures as a
function of density.
We have not explicitly included the results of the model of Lattimer and 
Swesty \cite{Lattimer:1991nc} in this comparison. 
They use the same excluded volume prescription as 
Shen et al.~\cite{Shen:1998gq} to take into account the medium dependence 
of the $\alpha$-particle fraction. 
Therefore, the behavior with density of their results is similar to that of 
Shen et al., even though absolutely there are differences due to a different 
treatment of the heavy particle fraction. 
A direct comparison of  $X_{\alpha}$ of Shen et al. and Lattimer et
al.\ \cite{Lattimer:1991nc}, can 
be found in Ref.\ \cite{Horowitz:2006pj}.

At low densities all models show a decreasing
$\alpha$-particle fraction with decreasing total baryon number density.
However, there are some subtle differences.
The virial expansion (black dash-dotted lines) takes scattering 
contributions to the second virial coefficient into account. 
These are important at higher temperatures and cause a slightly higher
prediction for $X_{\alpha}$ at small $n$.
The effect of cluster dissociation is absent in this model and results 
in an monotonic increase of the cluster fraction with increasing density, 
which is unphysical since one does not expect to see clusters in 
nuclear matter at saturation density. 

The NSE approach (green dotted lines) 
compared to the virial approach shows a reduction of the $\alpha$-particle 
fraction at all densities, due to the formation of heavier clusters.
$\alpha$-particles survive even beyond the nuclear saturation 
density, since medium modification and dissolution of clusters are absent 
in this model, which also fails to describe the transition from clusterized 
matter to cluster-free nuclear matter at high densities.

The Shen et al.\ EoS (blue dashed lines) neglects 
the contribution of deuteron, triton and helion clusters.
This leads to an overestimation of
the $\alpha$-particle fraction at low densities where 
actually lighter clusters dominate the composition.
In contrast to the above two approaches, there is a steep decrease of 
the $\alpha$-cluster abundance when approaching the saturation density.
However, one notes some irregularities which are understood as an effect of 
the excluded volume approach when the closest packing is reached.

The generalized RMF model (red solid lines) 
developed in this work describes 
the decrease of the $\alpha$-particle fraction at high densities by a 
reduction of their binding energy due to the Pauli blocking which leads to the 
Mott effect for vanishing binding.
The maximum cluster density is reached around the Mott density. 
Due to the presence of
strong correlations in the scattering state continuum
which are effectively  represented by one resonance,
there is a nonvanishing cluster fraction above the Mott density.
Among all models presented in the comparison of Fig.~\ref{fig:13}, the 
generalized RMF approach shows the strongest reduction of the 
$\alpha$-cluster fraction.
 
In the QS approach the behavior of the density dependence of 
$X_{\alpha}$ is similar to that of the generalized RMF model,
but has for all densities a higher $\alpha$-cluster fraction,
which is accompanied by smaller light cluster abundances.
As discussed in Subsect.\ \ref{subsec:comp} the difference in $X_{\alpha}$ 
between the two models is mainly due to the overemphasis of the deuteron
correlations in RMF, which suppresses the  $\alpha$-particle fraction. 
On the other hand, the RMF takes into account the back-reaction of the cluster 
formation on the 
mean-field which is missing in the QS approach.

The difference between the two approaches of this paper, 
shown in Fig.~\ref{fig:13} in direct comparison, also gives an indication 
of the effects of possible improvements in the models, 
apart, of course, from the inclusion of heavier clusters, resp. of nuclei 
embedded in a clusterized gas. On the other hand,
the synopsis of Fig.~\ref{fig:13} demonstrates the advantage 
of systematic many-body approaches to the description of cluster formation
over alternative approaches which lack a microphysical mechanism to account 
for the cluster breakup, but which are extensively used, 
e.g., in astrophysical applications. 

\section{Conclusions}
\label{sec:Concl}

Up to now there exist different strategies to 
model the EoS of nuclear matter  and, in particular, to extract the 
symmetry energy: 
(1) phenomenological density functional methods, such as non-relativistic 
Skyrme or relativistic RMF functionals, 
(2) effective-field theory approaches based on density functional or chiral 
perturbation theory, or 
(3) {\it ab initio} approaches, such as Brueckner-type methods, 
variational calculations or Green function methods. 
A recent overview over these methods with references can be found, e.g., in 
Ref. \cite{Fuchs:2005yn}.
These methods are developed to obtain reliable equations of state for
nuclear matter for a range of densities and asymmetries. 
In particular, the density-dependent RMF model
can be considered phenomenologically a very useful approach in this respect
since it has been applied with great success in the simultaneous description 
of cold nuclear matter in finite nuclei and compact stars as well as of hot
nuclear matter in heavy-ion collisions and supernova explosions.
However, all these approaches fail in the low-density limit, where cluster
formation becomes essential. 
In this region, simple approaches which take clusters into account,
such as the NSE or the Beth-Uhlenbeck formula of the 
virial expansion, show that clusters give a substantial 
contribution to the composition and the thermodynamic properties.
On the other hand, models like the NSE or the virial expansion
fail at higher densities where in-medium effects become important
leading to a dissolution of clusters and the transition to cluster-free
nuclear matter.

Here we propose, for the first time, a unified treatment
that takes both limits into account describing
the smooth transition from clusterized matter at low densities to
pure nucleonic matter at high densities.
We thus suggest a symbiotic framework which combines the merits 
of a QS approach in describing cluster properties in a medium 
with those of the RMF approach to model nucleonic self-energy effects, 
resulting in two hybrid approaches to the problem of cluster formation and 
dissolution.
On the one hand we use a microscopic QS approach to describe the 
medium modification of cluster binding energies
due to Pauli blocking. In this approach we employ nucleon self-energies
taken in parametrized form from a very recent version of the 
density-dependent RMF model. 
On the other hand, the results of
the QS approach for the cluster properties inside the medium are
incorporated into an effective 
hadronic field theory using a Lagrangian formulation.
Such a theory already has the correct high-density behavior 
as deduced from comparisons to heavy-ion collision experiments. 
We improve the low-density behavior by explicitly including light clusters, 
such as deuterons, tritons, helions
and $\alpha$-particles as explicit degrees of freedom
taking into account the medium modification of their binding energies
from the results of the QS approach.

In the numerical evaluation of both hybrid approaches developed 
in this work, we
find that well-defined clusters appear only for densities below 
approximately $1/10$ to $1/100$ of the saturation density 
and get dissolved at higher densities.
A direct confirmation of the given approach can be obtained from a comparison
with recent results from heavy-ion collisions at low-energies 
\cite{Kowalski:2006ju}.
These investigations indicate larger values for the symmetry energy in 
comparison with the mean field results at low densities, which 
seem to be in 
agreement with our findings \cite{progress}.

Realistic approaches to the clustering in low-density nuclear matter
should include excited states such as resonances 
and also the contribution of the continuum of scattering states.
This can be done for the second virial coefficient, as demonstrated by the 
generalized Beth-Uhlenbeck approach \cite{Schmidt:1990}.  
In this way one can also reach the exact low-density limit of the 
virial approach \cite{Horowitz:2006pj,Horowitz:2005nd,O'Connor:2007eb}. 
Compared with the approaches of Lattimer et al.~\cite{Lattimer:1991nc} 
and Shen et al.~\cite{Shen:1998gq}, who employ
phenomenological concepts such as the excluded volume, we have given
in this work a description of medium effects on the clusters and of their 
breakup as a result of the fundamental Pauli principle.

The extension of the present framework to larger clusters beyond the 
$\alpha$-particle is straightforward along the lines given 
in this work.
For the evaluation of self-energies and Pauli shifts of A-particle clusters in
nuclear matter, see \cite{Ropke:1984}.
The generalization of the given approach to account for clusters of arbitrary 
size would lead to an improvement 
in the low-density limit when comparing
the nuclear statistical equilibrium as used, e.g., in multifragmentation 
models \cite{Bondorf:1995ua,Gross:1990}.
One can alternatively also introduce the formation of heavier nuclei 
\cite{Ropke:1984} in the presence of a nucleon and cluster gas, in a similar 
way as it was done in the Thomas-Fermi approximation in the Shen et al.\ 
approach. This will be relegated to a subsequent paper. 
We restricted our present work to that region of the phase diagram 
where heavier clusters with $A > 4$ are not relevant. 

In our models, we constructed the
phase transition from clusterized, gaseous low-density matter to 
a cluster-free nuclear liquid at high densities. 
The coexistence region gives a hint about the range in temperature and density
where the occurrence of inhomogeneities and the formation of heavy
clusters become relevant.
An issue for future investigations is an improved description 
near the phase transition taking into account effects of the
Coulomb interaction and charge screening.

We are able to give the composition and the thermodynamic quantities in a 
large region of densities, temperatures and asymmetries as they are required, 
e.g., in supernova simulations.
We did not consider contributions from, e.g., electrons, neutrinos or
photons, to the thermodynamical quantities. In astrophysical
applications of the EoS, they have to be included. They will modify
the properties of the system and affect, in particular, the occurence
of inhomogeneities and of the liquid-gas phase transition.

As a long-range objective, we aim at a unified description of nuclear 
matter from very low density to, eventually, the deconfinement phase 
transition that is based on a more microscopic and self-contained
description than previous approaches to the EoS which are used up to now 
in astrophysical models.

\section*{Acknowledgements}

This research was supported by the DFG cluster of excellence ``Origin
and Structure of the Universe''
and by CompStar, a Research Networking Programme of the European 
Science Foundation.
The work of TK was supported by the Department of Energy, Office of Nuclear
Physics, contract no.\ DE-AC02-06CH11357. DB acknowledges support from the 
Polish Ministry for Research and Higher Education under grants 
No. N N202 0953 33, No. N N202 2318 37 and from the Russian Fund for 
Fundamental Investigations under grant No. 08-02-01003-a.

\begin{appendix}
\section{Low-density expansion}
\label{sec:app1}

A density-dependent RMF model was considered in \cite{Typel:2005ba}.
The following low-density expansions are derived form this model and reproduce 
the DD-RMF results below the baryon density $n \le 0.2$ fm$^{-3}$ 
within 0.1 \%; variables are the total baryon density
$n= n_{p}^{\rm tot}+n_{n}^{\rm tot}$ in units of fm$^{-3}$, the asymmetry 
$\delta = (n_{p}^{\rm tot}-n_{n}^{\rm tot})/(n_{p}^{\rm
  tot}+n_{n}^{\rm tot})$ and the temperature $T$ in MeV.
The scalar field in MeV is given by
\begin{eqnarray}
 \Sigma_{n,p}(T,n,\delta) &=& 
 n_B [(4524.13-6.926 \: T)-14.5157 \delta^{2}/4
 +0.833943 \: \delta^{4}/16 -9.00693 \: \delta^{6}/64] 
 \nonumber\\&&
 + n^2 [-19190.7-2426.57 \: \delta^{2}/4
     -317.732 \: \delta^{4}/16-1547.38 \:  \delta^{6}/64]
 \nonumber\\&& 
 + n^3 [62169.5+2521.29 \: \delta^{2}/4 + 3470.28 \: \delta^{4}/16] 
 \nonumber\\&& 
 + n^4 [-91005.1+3984.82 \: \delta^{2}/4-9148.6 \: \delta^{4}/16];
\end{eqnarray}
and the vector field in MeV by
\begin{eqnarray}
 \Sigma_{p}^{0}(T, n, \delta)&=& 
 \Sigma_{n}^{0}(T, n,- \delta) = 
 n [3462.24+946.705 \: \delta/2-0.334508 \: \delta^{2}/4] 
 \nonumber\\&& 
 + n^2 [-11312.4-6246.21 \: \delta/2 - 6353.53 \: \delta^{2}/4
 -0.099478 \: \delta^{3}/8]
 \nonumber\\&&
 + n^3 [20806.1 +18717.6 \: \delta/2  +29298. \: \delta^{2}/4
 -0.490543 \: \delta^{3}/8] 
 \nonumber\\&& 
 + n^4 [352.371-24887.2 \: \delta/2  -39807.4 \: \delta^{2}/4
 -0.346218 \: \delta^{3}/8].
\end{eqnarray}
The vector field is nearly independent of temperature, the scalar field has 
a weak temperature dependence.

\end{appendix}

\end{document}